\PassOptionsToPackage{hidelinks}{hyperref}

\documentclass[pdflatex,sn-mathphys-num]{sn-jnl}

\makeatletter
\@ifpackageloaded{geometry}{\geometry{reset}}{\RequirePackage{geometry}}
\makeatother

\usepackage[utf8]{inputenc}
\usepackage[T1]{fontenc}

\usepackage{libertine}            
\usepackage[varqu]{zi4}           

\usepackage{amsmath,amssymb,amsfonts}

\usepackage[libertine]{newtxmath}

\usepackage{bm}                   

\usepackage{graphicx}
\usepackage{multirow}
\usepackage{rotating}
\usepackage[title]{appendix}
\usepackage{xcolor}
\usepackage{textcomp}
\usepackage{manyfoot}
\usepackage{array}
\usepackage{booktabs}
\usepackage{tabularx}
\usepackage{adjustbox}
\usepackage{pifont}
\usepackage{tikz}

\usepackage{algorithm}
\usepackage{algorithmicx}
\usepackage{algpseudocode}

\usepackage{listings}

\graphicspath{{./Figures/}{./}}

\usepackage{enumitem}

\usepackage{siunitx}

\ExplSyntaxOn
\AtBeginDocument{
  \sisetup{
    mode                   = math,
    locale                 = US,
    output-decimal-marker  = {.},
    group-digits           = integer,
    group-minimum-digits   = 4,
    group-separator        = {\,},
    detect-weight          = true,
    detect-family          = true,
    table-number-alignment = center,
    input-decimal-markers  = {.},
    input-ignore           = {,}
  }
}
\ExplSyntaxOff

\newcolumntype{Y}{>{\centering\arraybackslash}X}

\newcolumntype{Z}[1]{S[
  table-format = #1,
  table-number-alignment = center
]}

\newcommand{\n}[1]{\num{#1}}
\newcommand{\pmin}{$<\!10^{-16}$}



\usepackage{hyperref}

\definecolor{byzantine}{rgb}{0.74, 0.2, 0.64}
\newcommand{\CHANGES}[1]{{#1}}

\definecolor{acceptedblue}{RGB}{232,244,255}
\definecolor{acceptedborder}{RGB}{46,125,50}

\newcommand{\AcceptedManuscriptNote}{%
\begingroup
\setlength{\fboxsep}{6pt}%
\setlength{\fboxrule}{0.5pt}%
\noindent\fcolorbox{acceptedborder}{acceptedblue}{%
\parbox{\dimexpr\textwidth-2\fboxsep-2\fboxrule\relax}{%
\normalfont\small
\textbf{NOTE:}
This article has been accepted for publication in the \textit{Journal on Wireless Communications and Networking}
(Springer Nature). It is an extended journal version of our conference paper:
``\href{https://doi.org/10.1109/EuCNC/6GSummit63408.2025.11037092}{A Statistical Evaluation of Indoor LoRaWAN Environment-Aware Propagation for 6G: MLR, ANOVA, and Residual Distribution Analysis},''
published in the \textit{2025 Joint European Conference on Networks and Communications \& 6G Summit
(EuCNC/6G Summit)}, pp.~494--499, June 2025.
}%
}%
\par
\endgroup
}

\begin{document}

\title[Environment-Aware Indoor LoRaWAN Path Loss]{%
\AcceptedManuscriptNote
\vspace{4em}

Environment-Aware Indoor LoRaWAN Path Loss: Parametric Regression Comparisons, Shadow Fading, and Calibrated Fade Margins%
}

\author*[1]{\fnm{Nahshon Mokua} \sur{Obiri}}\email{nahshon.obiri@student.uni‑siegen.de}
\author[1]{\fnm{Kristof} \sur{Van Laerhoven}}\email{kvl@eti.uni‑siegen.de}

\affil[1]{\orgdiv{Department of Electrical Engineering and Computer Science},
\orgname{University of Siegen},
\orgaddress{\country{Germany}}}

\abstract{Indoor long range wide area network (LoRaWAN) propagation is shaped by structural and time-varying environmental factors, which limit single-slope log-distance models and the standard log-normal shadowing assumption. We propose an environment-conditioned path loss framework that augments a log-distance multi-wall baseline with co-recorded environmental covariates (relative humidity, temperature, carbon dioxide, particulate matter, and barometric pressure) and receiver-reported signal-to-noise, and we validate both the mean and the residual law statistically. The approach is evaluated on a $12$-month campaign in an eighth-floor office ($240\,\mathrm{m}^2$) using time-blocked $5$-fold cross-validation and a chronological hold-out. Across parametric regressors (regularized multiple linear regression (MLR), conjugate Bayesian linear regression, and a selective quadratic MLR extension on continuous predictors), the selective polynomial mean improves out-of-sample accuracy, reducing cross-validated root mean square error from $8.23$ to $7.38\,\mathrm{dB}$ and increasing $R^2$ from $0.81$ to $0.84$. Out-of-fold (OOF) residuals are distinctly non-Gaussian and are best summarized by a compact $3$-component Gaussian mixture with a sharp core and a light, broad tail. Finally, we translate prediction error into reliability by prescribing the fade margin as the upper-tail percentile of OOF errors, attaching moving-block bootstrap uncertainty, and validating the resulting outage on a held-out set. At a $1\%$ outage target ($99\%$ reliability), the polynomial model requires $25.73\,\mathrm{dB}$ versus $27.79$ to $28.05\,\mathrm{dB}$ for linear baselines, enabling tighter indoor massive Internet of Things link budgets aligned with sixth-generation reliability targets under energy constraints.}


\maketitle
\section{Introduction}
\label{sec:intro}

Indoor long range wide area network (LoRaWAN) technology deployments are challenging to model because propagation is shaped by building materials and layout, time-varying occupancy, and environmental dynamics such as humidity and temperature \cite{cattaniExperimentalEvaluationReliability2017}. These factors induce rich multipath, attenuation discontinuities, and shadowing that break stationarity and yield heavy-tailed, heteroskedastic errors, violating the assumptions of classical log-distance path loss models (LDPLMs) and leading to underestimated uncertainty and fade margins \cite{grubelDenseIndoorSensor2022b}. Moreover, statistical validation criteria must be met when developing reliable path loss models. Specifically, it is essential to assess the statistical significance of model parameters, such as the path loss exponent, using analysis of variance (ANOVA). \CHANGES{Furthermore, many classical formulations commonly assume that the residuals (shadow fading)} \textbf{(i)} follow an approximately Gaussian law in decibels ($\mathrm{dB}$) (i.e., log-normal in linear power), \textbf{(ii)} exhibit approximately constant variance (homoskedasticity), and \textbf{(iii)} are approximately independent of the predictors (e.g., distance, structural composition, environmental parameters) and weakly autocorrelated \cite{gonzalez-palacioLoRaWANPathLoss2023}. \CHANGES{In this work, we test these assumptions diagnostically rather than presupposing them.}

Motivated by the stringent reliability and adaptability requirements envisioned for sixth-generation (6 G) wireless networks, such as ultra-reliable and low-latency communication (URLLC) and massive machine-type communication (mMTC) \cite{siddikyComprehensiveExploration6G2025}, this work extends conventional indoor LoRaWAN propagation modeling. We complement recent outdoor, weather‑driven LoRaWAN predictability studies, such as \cite{gonzalez-palacioLoRaWANPathLoss2023} and \cite{szafranskiPredictabilityLoRaWANLink2024}, by bringing the same idea inside the building. We propose and validate an advanced statistical modeling approach that explicitly integrates environmental variables into parametric regression techniques, ensuring robust statistical diagnostics. Specifically, we build upon our previous empirical characterization of an extensive dataset \cite{mokuaobiriComprehensiveDataDescription2025}, comprising a $12$-month measurement dataset collected on the 8th floor of a single $ 240\,\mathrm{m}^2$ operational office space at $ 250\,\mathrm{m}$ above sea level, as elaborated in Table \ref{tab:RQS} of research questions (RQs).

\begin{table}[hbt]
    \centering
    \caption{\textbf{Research questions guiding this study}}
    \label{tab:RQS}
    \begin{tabular*}{\columnwidth}{@{\extracolsep{\fill}} >{\centering\arraybackslash}p{0.08\linewidth} p{0.88\linewidth} @{}}
        \toprule
        \textbf{\#} & \textbf{Research Question} \\
        \midrule
        RQ1 & Do environment-aware covariates deliver statistically significant and practically meaningful improvements over structure-only baselines, as verified by heteroskedasticity-robust \textit{Type}~II/III analysis of variance (ANOVA) and nested partial-$F$ tests? \\
        \midrule
        RQ2 & Which parametric mean model among regularized multiple linear regression (MLR), conjugate Bayesian linear regression (BLR), and a selective quadratic extension on continuous predictors for the MLR, achieves the best bias-variance trade-off under time-ordered $k$-fold cross-validation? \\
        \midrule
        RQ3 & What distribution best characterizes out-of-fold (OOF) shadow fading among single-component parametric baselines and Gaussian mixture models (GMMs), according to the goodness-of-fit criteria and residual diagnostics? \\
        \midrule
        RQ4 & How do the residual upper-tail quantiles map to calibrated fade margins at target outages of  $1$–$5\%$, and what savings in $\mathrm{dB}$ do they enable relative to fixed margin heuristics? \\
        \bottomrule
    \end{tabular*}
\end{table}

We extend our prior study \cite{obiriStatisticalEvaluationIndoor2025a} by moving beyond multiple linear regression (MLR) to a comparative analysis of three parametric families: \textbf{(i)} baseline MLR with ridge regression (least squares with an $\ell_{2}$ (squared Euclidean norm) penalty, also known as Tikhonov regularization), least absolute shrinkage and selection operator (Lasso) with an $\ell_{1}$ (absolute-value / sparsity) penalty, and elastic net regularization; \textbf{(ii)} POLY2, a selective quadratic extension MLR that adds second‑order terms only for continuous predictors (distance, environment, the signal-to-noise ratio (SNR)), while keeping wall-count terms linear, and \textbf{(iii)} Bayesian linear regression (BLR) with conjugate priors, namely the normal inverse gamma prior and the Zellner $g$ prior. This comparison addresses nonlinear propagation behavior and multicollinearity among environmental predictors (discussed in Sec.~\ref{subsec:statsregression}). Under a $5$-fold cross-validation, the second-order polynomial emerged as superior, confirming nonlinear dependencies between environmental factors and signal attenuation and substantially reducing prediction errors relative to the linear alternatives.

Integrating environmental sensing directly into the path loss modeling process is crucial for deploying context-aware, resilient indoor Internet of Things (IoT) infrastructures, as targeted by 6 G \cite{siddikyComprehensiveExploration6G2025}. It enables accurate network planning, adaptive resource allocation, and power optimization in dynamic scenarios where environmental and occupancy conditions continuously fluctuate. As observed in \cite{gonzalez-palacioMachineLearningBasedCombinedPath2023}, while continuous monitoring of these parameters adds moderate complexity and power overhead, deploying low-cost environmental sensors alongside LoRaWAN devices provides actionable, real-time insights that enable dynamic adjustment of path loss predictions, precise calibration of fade margins accounting for residual multimodality, and consequently, robust and reliable network performance.

Our analytical framework incorporates a detailed statistical validation procedure that combines comparative parametric regression with classical predictors of distance and wall counts, along with weights for five environmental factors: relative humidity, temperature, CO\textsubscript{2}, particulate matter, and barometric pressure. We apply heteroskedasticity-robust \textit{Type}~II and \textit{Type}~III ANOVA, together with nested partial $F$ tests, to validate predictor relevance and interactions and to quantify the reduction in unexplained variance attributable to the environmental terms. \CHANGES{Shadow fading is examined using parametric distributions, including Normal, Skew--Normal, and GMMs, as well as nonparametric methods such as kernel density estimation (KDE), bootstrapping for confidence interval (CI) estimation, and Kruskal--Wallis tests for group-wise location shifts, and Brown--Forsythe/Levene-type tests for group-wise dispersion (variance heterogeneity)}. Finally, we treat fade margin as a reliability-controlled buffer inferred from the upper tail of cross-validated residuals, and we verify calibration on held-out data via the achieved outage and its complement, the achieved reliability (non-exceedance) on received packets. This calibration operationalizes our statistical gains into a deployable reliability control parameter (fade margin) for network planning with minimal overhead.

This integrated methodological toolkit provides deeper theoretical insights and directly addresses the practical deployment challenges envisioned for 6 G wireless systems. The outcomes of this research provide a framework for future investigations, including machine learning (ML) extensions, multi-site validations, and real-time adaptive propagation models, thereby contributing to the development of reliable, sustainable, and context-aware indoor IoT networks. The contributions of this work are:

\begin{enumerate}[label=(\roman*)]
    \item \CHANGES{A head-to-head statistical evaluation of physically grounded parametric regressors using time-blocked $5$-fold cross-validation, including regularized linear MLR,  conjugate BLR with shrinkage priors, and regularized selective second-order MLR. The best second-order model reduced the root mean square error (RMSE) from $8.23$ to $7.38\,\mathrm{dB}$ and increased $R^2$ from $0.81$ to $0.84$ (Table~\ref{tab:perf_comparison}).}
    \item Explicit, heteroskedasticity-robust \textit{Type}~II/III ANOVA with nested partial-$F$ tests that quantify predictor relevance. Relative to a structure-only baseline, adding environment-aware covariates and SNR reduces unexplained variance by $44\%$ (Section~\ref{sec:anova}, Table~\ref{tab:mlr_significance}).
    \item A residual (shadow-fading) characterization from OOF errors. Goodness-of-fit criteria and diagnostics favor a $3$-component Gaussian mixture model (GMM) over single-component baselines. KDE and groupwise dispersion tests confirm a sharp core with a light, broad tail (Table~\ref{tab:distfits}).
    \item A quantile-based fade margin rule, $\widehat{M}_{\mathrm{F}}(p)=Q_{1-p}(r)$, \CHANGES{with $p$ as the target outage probability,} with bias-corrected and accelerated (BCa) moving-block bootstrap uncertainty and held-out validation. At $p=1\%$, the selective polynomial mean requires $25.73\,\mathrm{dB}$ versus $27.79$ to $28.05\,\mathrm{dB}$ for linear baselines ($\gtrsim 2.2\,\mathrm{dB}$ lower; Eq.~\eqref{eq:fmp}, Table~\ref{tab:fm_calibration}, Fig.~\ref{fig:rho_vs_fm}).
\end{enumerate}

The remainder of this paper is organized as follows: Section~\ref{sec:background} reviews indoor LoRaWAN propagation and motivates the incorporation of environmental context. Section~\ref{sec:methodology} details the experimental setup, feature construction, model specification, validation protocol, and fade margin calibration procedure. Section~\ref{sec:results} reports comparative predictive performance and ANOVA effect sizes, characterizes the residual (shadow fading) distribution, and translates residual upper-tail behavior into calibrated fade margins validated on a held-out set. \CHANGES{Section~\ref{sec:conclusion} concludes with the main takeaways, and Section~\ref{sec:future} outlines limitations and future directions.}

\section{Background}\label{sec:background}

6 G roadmaps consistently foreground three constraints for massive IoT: \textbf{(i)} reliability under nonstationary conditions, \textbf{(ii)} energy efficiency within strict duty-cycle and battery limits, and \textbf{(iii)} context awareness so that networks sense and adapt to their environment rather than treating it as exogenous noise \cite{siddikyComprehensiveExploration6G2025}. In this setting, LoRaWAN plays a complementary role to higher-throughput 6 G interfaces by offering sub-GHz penetration and multi-year device lifetimes for dense indoor sensing, as we established in our recent survey \cite{obiriSurveyLoRaWANIntegratedWearable2024}, provided that propagation models are site-specific and time-aware to support low-margin, reliable links. Conditioning path loss and shadowing on ambient state reduces unexplained variance and enables tighter fade margins with lower transmit energy in practice \cite{gonzalez-palacioMachineLearningBasedCombinedPath2023}. Our 6 G-aligned statistical evaluation similarly showed that incorporating environmental covariates reduces unexplained variance and reveals multimodal residual structure that standard single-slope baselines miss \cite{obiriStatisticalEvaluationIndoor2025a}. \CHANGES{ Accordingly, this section connects these 6 G constraints to the case for environment-conditioned indoor LoRaWAN propagation models adopted in this work.}

\subsection{Indoor LoRaWAN Signal Propagation}

Indoor LoRaWAN propagation is notoriously complex due to the rich scattering and attenuation effects caused by walls, floors, and clutter in buildings \cite{azevedoCriticalReviewPropagation2024}. Most empirical wireless network propagation models, including LoRaWAN, assume a logarithmic increase in path loss with distance, as per the LDPLM. This simple model often serves as a valuable starting point, but it does not capture all the complexities of indoor environments. In practice, measured path loss exponents in buildings vary widely.  Corridor measurement in \cite{robles-encisoLoRaZigbee5G2023a} found an apparent path loss slope below $2.0$ in one non-line-of-sight (NLoS) region and above $3.0$ in a more obstructed section. Such variability reflects how indoor layout and materials either guide propagation, for example, along corridor waveguides, or impose excess loss through multiple brick walls. For instance, campus-scale studies in \cite{anisahExperimentalResultsLoRa2023} report exponents from $1.37$ (indoor line-of-sight (LoS)) to values greater than $2.3$ (indoor NLoS), highlighting that internal materials and layout dominate performance and that one-size exponents are rarely adequate. Moreover, it has also been observed in \cite{bertoldoEmpiricalIndoorPropagation2019, anisahExperimentalResultsLoRa2023} that uncalibrated models tend to either over- or under-predict coverage.

To address structural variability, refined models explicitly encode building features. Comparative evaluations against indoor LoRa measurements (e.g., ITU-R P.1238 \cite{RECOMMENDATIONITURP123811}, COST 231 multi-wall model (MWM) \cite{europeancommissionCOSTAction2311999}, Motley–Keenan \cite{limaMotleyKeenanModelAdjusted2005}) show that multi-wall formulations often deliver the best accuracy when parameters are locally calibrated \cite{bertoldoEmpiricalIndoorPropagation2019}. However, even with careful calibration, purely structural models fail to capture temporal influences. For example, multi-floor residential measurements in \cite{zhongMeasurementModelingLoRa2025} show pronounced room- and time-dependent fading (\CHANGES{standard deviation} up to $18.4\,\mathrm{dB}$ in LoS), pointing to occupancy and heating, ventilation, and air conditioning (HVAC) cycles as additional covariates of variability.

Consequently, recent work has shifted toward hybrid or semi-empirical approaches, starting from a physically anchored path loss law, then learning data-driven corrections that reflect site materials, geometry, and temporal context \cite{alkhazmiAnalysisRealWorldLoRaWAN2023, azevedoCriticalReviewPropagation2024}. This motivates our stance of augmenting structural predictors with environmental variables (e.g., humidity, temperature, CO$_2$, pressure, particulate matter) and validating them using intensive statistics rather than assuming log-normal shadowing and homoscedasticity by default. We now operationalize this by comparing parametric regressors using cross-validation, quantifying effect sizes via ANOVA, and characterizing shadow fading with mixture models, thereby aligning with background theory and providing a deployment-ready, reliability-calibrated model for indoor LoRaWAN propagation.

\subsection{Environment-aware LoRaWAN Propagation}

Beyond static architecture, indoor links are shaped by time-varying context: occupancy, HVAC cycles, and the microclimate. Long-duration deployments show that temperature and humidity fluctuations correlate with shadow fading over days \cite{muppalaInvestigationIndoorLoRaWAN2021}. On the other hand, crowded periods introduce human-body absorption and moving scatterers, increasing the variance of the received signal strength indicator (RSSI) and reducing short-term stability \cite{grubelDenseIndoorSensor2022b,harindaPerformanceLiveMultiGateway2022}. Even in office settings, Wi-Fi or Bluetooth Low Energy (BLE) links become erratic during busy hours and stabilize overnight \cite{sadowskiRSSIBasedIndoorLocalization2018}. These observations suggest that purely structural models cannot capture nonstationarity.

Outdoor long‑term evidence shows diurnal weather structure in SNR and link‑specific sensitivity to temperature and humidity \cite{szafranskiPredictabilityLoRaWANLink2024}. From a physics perspective, microclimate mechanisms explain the measurable path loss drift. For instance, humidity modifies the effective permittivity of air and porous materials while temperature and air-density changes alter boundary conditions, which shape attenuation at sub-GHz \cite{gonzalez-palacioLoRaWANPathLoss2023,guerraForecastingLoRaWANRSSI2024}. Empirically, higher indoor temperatures or associated ventilation states can coincide with slightly lower path loss, while elevated CO\textsubscript{2} tracks occupancy and richer multipath \cite{obiriStatisticalEvaluationIndoor2025a}. Thus, variables such as temperature, relative humidity, CO\textsubscript{2}, pressure, and particulate matter serve as practical proxies for the environment's latent state, which governs propagation.

Motivated by this, recent work augments path loss models with co-located environmental sensing. In our prior study \cite{mokuaobiriComprehensiveDataDescription2025}, injecting CO\textsubscript{2}, humidity, temperature, pressure, and particulate matter into an LDPLM and wall-loss baseline reduced unexplained variance by over $40\%$, because the model could condition on occupancy and ventilation rather than treating their effects as noise. Others similarly report gains when connectivity control becomes environmental-aware \cite{gonzalez-palacioLoRaWANPathLoss2023}. Therefore, environment-aware modeling, which conditions on occupancy and ventilation via co-located sensing, sharpens link budgets and makes fade margins situational rather than static, yielding more reliable indoor LoRaWAN deployments.

Concurrently, the field is moving beyond curve fitting toward statistically rigorous and adaptive models. Hybrid methods add learned corrections to physics-grounded baselines \cite{aksoyComparativeAnalysisEnd2024}, and dynamic filters improve distance estimates by tempering fast fading and transient blockages \cite{voAdvancePathLoss2024}. In this paper, we take a principled approach: we compare parametric regressors that balance interpretability and nonlinearity, validate our contributions with ANOVA, and investigate shadow fading using mixture models and nonparametric diagnostics. The result is an environment-aware formulation grounded in testable assumptions and a reliability-calibrated interface for network planning via quantile-based fade margins.

\subsection{Statistical Regression Methods}
\label{subsec:statsregression}

Modeling indoor LoRaWAN propagation spans from transparent linear formulations to more flexible parametric and ML approaches. A recurring theme is the trade-off between interpretability, needed to attribute loss to distance, walls, or climate, and flexibility to capture mild nonlinearities and correlated predictors. Accordingly, studies commonly adopt linear baselines for physical clarity, then introduce controlled extensions, such as polynomial terms or regularization, and assess adequacy using cross-validation, information criteria, and residual diagnostics to ensure that out-of-sample gains warrant the added complexity. This section reviews representative linear baselines and polynomial extensions.

\subsubsection{Multiple Linear Regression}
MLR is a foundational regression technique that models path loss as a linear combination of multiple predictors, including distance, wall penetration, and environmental parameters (e.g., humidity, temperature). \CHANGES{If $V$ denotes the number of predictors,} it takes the general form in Eq.~\eqref{eq:mlr}:
{\small
\begin{equation}
    \label{eq:mlr}
    y = \beta_0 + \beta_1 x_1 + \beta_2 x_2 + \cdots + \beta_V x_V + \psi,
\end{equation}
}
where $y$ represents the observed path loss, $\beta_0$ is the intercept, $\beta_1,\ldots,\beta_V$ denote regression coefficients, $x_1,\ldots,x_V$ are the predictors, and $\psi$ is the residual error \cite{farawayPracticalRegressionAnova2002}, typically assumed to be normally distributed ($\psi \sim \mathcal{N}(0,\sigma^2)$, \CHANGES{where $\sigma^2$ is the residual (noise) variance}).

A primary advantage of MLR is its interpretability. Each coefficient corresponds to a physically meaningful impact, such as attenuation per wall or unit distance, making it particularly valuable for practical engineering applications and network planning. Beyond estimating these coefficients, an ANOVA can be performed to confirm the overall statistical significance of the regression and to determine whether each predictor makes a meaningful contribution. For example, Bertoldo et al.\ \cite{bertoldoEmpiricalIndoorPropagation2019} used MLR to quantify propagation losses tied to specific building materials and structures. However, reliability depends on assumptions of linearity, approximate normality, homoscedasticity of residuals, and weak autocorrelation. Empirical work therefore emphasizes residual diagnostics, Quantile--Quantile plots, and tests such as Jarque--Bera (normality), Breusch--Pagan (heteroskedasticity), and Durbin--Watson (autocorrelation), to verify these assumptions \cite{gonzalez-palacioMachineLearningBasedCombinedPath2023}. Violations typically motivate carefully constrained extensions (polynomial terms or regularization) rather than abandoning the physically anchored linear form.

\subsubsection{Polynomial Regression}

Polynomial regression extends the capabilities of MLR by incorporating polynomial or interaction terms into the model to capture subtle nonlinearities commonly observed in indoor propagation environments. Let $V$ denote the number of predictors. Equation~\eqref{eq:poly2} gives a common second-order specification.
{\small
\begin{equation}
\label{eq:poly2}
    y = \beta_0 + \sum_{i=1}^{V}\beta_i x_i + \sum_{i=1}^{V}\sum_{j=i}^{V}\beta_{ij} x_i x_j + \psi,
\end{equation}
}

\CHANGES{where $\beta_{ij}$ are coefficients for second-order terms, yielding quadratic predictors when $i=j$ (i.e., $x_i^2$) and cross-product interaction predictors when $i<j$ (i.e., $x_i x_j$), with indices $i,j \in \{1,\ldots,V\}$. The constraint $j \ge i$ avoids duplicating symmetric interaction terms (since $x_i x_j = x_j x_i$), enabling the model to capture curvature and pairwise interactions between variables (e.g., distance--humidity coupling).} This extension is practical when the log-distance relationship departs slightly from strict linearity, since even a single quadratic term for distance can approximate multi-slope behavior and reduce error while preserving interpretability \cite{grubelDenseIndoorSensor2022b}. Studies typically select second-order terms via cross-validation or information criteria (Akaike information criterion (AIC) \cite{akaikeNewLookStatistical1974} or Bayesian information criterion (BIC) \cite{schwarzEstimatingDimensionModel1978}), and re-check residual assumptions (normality, homoscedasticity, and independence) because added terms may increase complexity and multicollinearity. In summary, polynomial regression is a compact extension that maintains interpretability while improving fit over MLR.

\subsubsection{Regression Regularization Methods}

Regularization enhances regression by introducing penalty terms that regulate variance and improve generalization in the presence of multicollinearity and limited sample sizes. These conditions are common in indoor LoRaWAN studies where environmental variables often co-vary. The three standard approaches are Ridge (an $\ell_2$ penalty), Lasso (an $\ell_1$ penalty), and Elastic Net (a mixture of $\ell_1$ and $\ell_2$) \cite{geronHandsonMachineLearning}. They differ in the penalty imposed and, consequently, in how coefficients are shrunk and selected.

\CHANGES{Here, $\mathbb{R}$ denotes the real numbers. Let $N$ denote the number of observations (samples) and let $V$ denote the number of predictors after feature mapping. Let $\mathbf{y}\in\mathbb{R}^{N}$ collect the observed path loss values, and let $\mathbf{1}\in\mathbb{R}^{N}$ denote the all-ones vector (so the intercept is not penalized). Let $\beta_0\in\mathbb{R}$ denote the intercept and let $\boldsymbol{\beta}\in\mathbb{R}^{V}$ denote the regression coefficient vector. Let $\boldsymbol{\Phi}\in\mathbb{R}^{N\times V}$ denote the standardized design matrix obtained from the physics-guided second-order feature map (nonlinear predictors included), and let $\mathbf{X}$ denote the linear design when no polynomial expansion is used.} Estimators are defined as solutions to penalized least-squares problems. The notation $\|\cdot\|_2^2$ denotes the Euclidean norm and $\|\cdot\|_1$ the $\ell_1$ norm; thus $\tfrac{1}{2N}\|\cdot\|_2^2$ is the per-sample average of squared residuals, and the factor $\tfrac{1}{2}$ is conventional and does not affect the minimizer. For ordinary MLR without polynomial terms, the feature map is the identity, hence $\boldsymbol{\Phi}\equiv\mathbf{X}$ and the problems below reduce to standard MLR. For Elastic Net (ENet), $\alpha\in[0,1]$ controls the $\ell_1/\ell_2$ mix ($\alpha=1$ yields Lasso; $\alpha=0$ yields Ridge up to the conventional scaling of $\lambda$ under this normalization) and $\lambda>0$ is the overall penalty strength. Equivalently, the two-parameter form $\lambda_1\|\boldsymbol{\beta}\|_1+\lambda_2\|\boldsymbol{\beta}\|_2^2$ corresponds to $\lambda_1=\lambda\alpha$ and $\lambda_2=\lambda(1-\alpha)/2$ under this normalization. Then, let $\mathbf{y}\in\mathbb{R}^N$ denote the path loss observations and let 
$\boldsymbol{\Phi}\in\mathbb{R}^{N\times V}$ be the standardized design matrix (intercept excluded).
Ridge, Lasso, and Elastic Net estimate $(\beta_0,\boldsymbol{\beta})$ by minimizing the
mean squared error plus an $\ell_2$ and/or $\ell_1$ penalty, with $\lambda>0$ controlling
overall shrinkage and $\alpha\in[0,1]$ controlling the $\ell_1/\ell_2$ mix.

{\small
\begin{equation}
\label{eq:ols-phi}
(\hat{\beta}_0,\hat{\boldsymbol{\beta}}_{\text{ols}}) \;=\;
\arg\min_{\beta_0,\boldsymbol{\beta}}
\;\frac{1}{2N}\left\|\mathbf{y}-\beta_0\mathbf{1}-\boldsymbol{\Phi}\boldsymbol{\beta}\right\|_2^2 ,
\end{equation}
}
{\small
\begin{equation}
\label{eq:ridge-phi}
(\hat{\beta}_0,\hat{\boldsymbol{\beta}}_{\text{ridge}}) \;=\;
\arg\min_{\beta_0,\boldsymbol{\beta}}
\;\frac{1}{2N}\left\|\mathbf{y}-\beta_0\mathbf{1}-\boldsymbol{\Phi}\boldsymbol{\beta}\right\|_2^2
\;+\; \frac{\lambda}{2} \left\|\boldsymbol{\beta}\right\|_2^2 ,
\end{equation}
}
{\small
\begin{equation}
\label{eq:lasso-phi}
(\hat{\beta}_0,\hat{\boldsymbol{\beta}}_{\text{lasso}}) \;=\;
\arg\min_{\beta_0,\boldsymbol{\beta}}
\;\frac{1}{2N}\left\|\mathbf{y}-\beta_0\mathbf{1}-\boldsymbol{\Phi}\boldsymbol{\beta}\right\|_2^2
\;+\; \lambda \left\|\boldsymbol{\beta}\right\|_1 ,
\end{equation}
}
{\small
\begin{equation}
\label{eq:enet-phi}
(\hat{\beta}_0,\hat{\boldsymbol{\beta}}_{\text{ENet}}) \;=\;
\arg\min_{\beta_0,\boldsymbol{\beta}}
\;\frac{1}{2N}\left\|\mathbf{y}-\beta_0\mathbf{1}-\boldsymbol{\Phi}\boldsymbol{\beta}\right\|_2^2
\;+\; \lambda \!\left(\frac{1-\alpha}{2}\left\|\boldsymbol{\beta}\right\|_2^2 + \alpha \left\|\boldsymbol{\beta}\right\|_1\right),
\quad \alpha\in[0,1].
\end{equation}
}

The standard ordinary least squares (OLS) objective in Eq.~\eqref{eq:ols-phi} is augmented with penalty terms to improve stability, interpretability, and out-of-sample performance. Lasso, defined by Eq.~\eqref{eq:lasso-phi}, uses an $\ell_{1}$ penalty that can drive some coefficients to zero and produce a compact model. In indoor Wi-Fi fingerprinting, such sparsity reduced localization error and, with a modest extension, enabled the identification of outlier access points \cite{khalajmehrabadiJointIndoorWLAN2017}. Ridge, given in Eq.~\eqref{eq:ridge-phi}, uses an $\ell_2$ penalty that shrinks all coefficients while keeping them nonzero, which is valuable when predictors are highly correlated. Comparative LoRa measurements report similar indoor behavior for Ridge and Lasso.
In contrast, simple linear models degrade outdoors, suggesting the importance of feature design and mild nonlinearity over the specific penalty choice in that scenario \cite{bhavanamExploringLoRaSignal2024}. Elastic Net, specified in Eq.~\eqref{eq:enet-phi}, combines $\ell_1$ and $\ell_2$ penalties and performs feature selection while retaining groups of correlated variables. It has demonstrated strong performance with correlated received signal strength features in a wireless local area network \cite{khalajmehrabadiJointIndoorWLAN2017} and has served effectively as a global regressor in visible-light positioning, remaining accurate with few training samples when paired with a lightweight residual-correction step \cite{linIndoorNLOSVLPSystem2025}. Across these works, predictors are standardized, and penalty parameters are selected using cross-validation. Elastic Net offers a balanced default for strongly correlated feature sets, which are common in indoor sensing. Ridge preserves stability without removing variables, whereas Lasso yields a compact, interpretable subset of predictors.

\subsubsection{Bayesian Linear Regression}

BLR models the relationship between path loss and predictors in a fully probabilistic manner, treating both the coefficients and the noise variance as random. \CHANGES{With a Gaussian likelihood and the conjugate Normal--Inverse-Gamma (NIG) prior over $(\boldsymbol{\beta},\sigma^2)$, Eq.~\eqref{eq:blr_nig} specifies the BLR model:}
{\small
\begin{equation}
\label{eq:blr_nig}
\mathbf{y} \mid \boldsymbol{\beta},\sigma^2 \sim \mathcal{N}\!\bigl(\beta_0\mathbf{1}+\mathbf{X}\boldsymbol{\beta},\;\sigma^2 \mathbf{I}_N\bigr),\quad 
\boldsymbol{\beta}\mid\sigma^2 \sim \mathcal{N}(\boldsymbol{\beta}_0,\sigma^2 \Sigma_0),\quad 
\sigma^2 \sim \mathrm{Inv\text{-}Gamma}(a_0,b_0),
\end{equation}
}
\CHANGES{$\mathbf{X}\in\mathbb{R}^{N\times V}$ stacks standardized predictors (e.g., log-distance, wall counts, humidity, temperature, CO\textsubscript{2}, pressure, particulate matter); $\sigma^2>0$ is the noise (residual) variance and $\sigma>0$ is the corresponding residual standard deviation; $\mathbf{I}_N$ is the $N\times N$ identity; $(\boldsymbol{\beta}_0,\Sigma_0)$ and $(a_0,b_0)$ are prior mean/covariance and shape/scale hyperparameters, with $\Sigma_0\in\mathbb{R}^{V\times V}$ and $\Sigma_0\!\succ\!0$ (positive definite).} Standardizing predictors (zero mean, unit variance) makes coefficient priors comparable across features and stabilizes inference. The posterior updates in closed form, and the posterior predictive distribution is Student-$t$, a practical benefit when residuals exhibit heavier tails than Gaussian \cite{liangMixturesPriorsBayesian2008}.

Regularization appears naturally as a prior choice where a zero-mean Gaussian prior over coefficients yields ridge-like shrinkage (i.e., Maximum A Posteriori (MAP) estimation with a Gaussian prior), whereas a Laplace prior corresponds to Lasso-type sparsity. Both provide stability when environmental covariates are correlated, without abandoning interpretability \cite{gelmanBayesianDataAnalysis2013}. A widely used alternative is Zellner’s $g$-prior, which ties the coefficient prior covariance to the observed design and preserves conjugacy (including in common mixtures of $g$ variants), providing analytic tractability and adaptive shrinkage in regression settings \cite{liangMixturesPriorsBayesian2008}. In particular, Eq.~\eqref{eq:zell} defines the prior:
{\small
\begin{equation}
\boldsymbol{\beta}\mid\sigma^2 \sim
\mathcal{N}\!\bigl(\mathbf{0},\,g\,\sigma^2(\mathbf{X}^\top \mathbf{X})^{-1}\bigr),
\label{eq:zell}
\end{equation}
}
\CHANGES{where $g>0$ is a scalar hyperparameter that controls the shrinkage strength and $(\mathbf{X}^\top\mathbf{X})^{-1}$ exists under full column rank, i.e., invertible. Here, $\mathbf{X}$ excludes the intercept column $\mathbf{1}$, while $\beta_0$ is treated separately and is not assigned a $g$-prior.}

Model assessment in modern Bayesian workflows typically relies on out-of-sample criteria computed from the full posterior distribution. Leave-one-out cross-validation (LOO) and the Widely Applicable Information Criterion (WAIC) are widely recommended. On the other hand, Pareto-smoothed importance sampling (PSIS-LOO) makes LOO efficient and diagnostically transparent (via shape-parameter checks), helping guard against overconfident fits \cite{vehtariPracticalBayesianModel2017}. When sparsity or group-wise shrinkage is desired (e.g., many correlated climate features), hierarchical priors such as the horseshoe concentrate mass near zero while preserving heavy tails for truly non-negligible effects, offering adaptive regularization without manual feature pruning.

\subsubsection{Machine Learning Regressors}

Beyond parametric approaches, indoor propagation studies increasingly employ ML models that learn nonlinear mappings from measurements to path loss with minimal functional assumptions \cite{elmezughiComparativeAnalysisMajor2022}. Commonly used models include Support Vector Regression (SVR), Gaussian Process Regression (GPR), tree ensembles (random forests and boosted ensembles), and neural networks. Kernel SVR captures moderate nonlinearities through flexible similarity measures while retaining a convex training objective. Its performance depends on the choice of kernel and the SVR hyperparameters $(C,\varepsilon)$, \CHANGES{where $C>0$ is the penalty (regularization) strength controlling the trade-off between model flatness and training errors, and $\varepsilon>0$ is the width of the $\varepsilon$-insensitive tube in the SVR loss.} GPR provides a probabilistic surrogate with uncertainty quantification via kernel covariances, but its $\mathcal{O}(N^3)$ training complexity in the number of samples $N$ often requires sparse or inducing-point approximations in long campaigns. Tree ensembles are robust to heterogeneous features and monotone transformations, handle interactions without requiring polynomial terms, and provide variable-importance profiles that facilitate interpretation. Neural networks (from shallow multilayer perceptrons (MLPs) to temporal convolutional neural networks (CNNs) or long short-term memory (LSTM) variants) can further reduce error when large and diverse datasets are available and when nonlinear couplings (e.g., distance$\times$occupancy proxies) are strong. However, they typically trade interpretability for accuracy and require careful regularization.

Reported performance for indoor LoRaWAN propagation is heterogeneous, with ML gains ranging from negligible to substantial, depending on the residual complexity remaining after accounting for dominant structural drivers and the available training data. When distance and obstruction proxies (e.g., wall counts) already capture much of the large-scale attenuation, incremental gains from purely data-driven models can be modest, and limited data further increases the risk of overfitting. Consequently, ML is frequently deployed in a complementary role, for example, as a correction layer on top of a physically grounded baseline \cite{hosseinzadehNeuralNetworkPropagation2017} or as a context-aware component that adapts to occupancy and microclimate indicators. Best practices emphasize leakage-safe evaluation (e.g., grouped and time-aware splits, nested validation, and stability checks) so that improvements reflect genuine generalization rather than shortcut learning. Overall, ML regressors are most effective when used in hybrid strategies that balance physical interpretability with flexible, data-driven corrections.

\subsection{Predictor Statistical Significance}
\label{sec:pred_sig}

A central question in indoor propagation studies is whether each predictor contributes uniquely to explaining path loss beyond correlated alternatives. Standard practice combines coefficient $t$-tests with partial-$F$ tests via ANOVA to assess main effects and interactions under a linear modeling framework \cite{montgomeryDesignAnalysisExperiments2017}. For nested models $\mathcal{M}_0 \subset \mathcal{M}_1$, the partial-$F$ statistic compares the residual sum of squares, \CHANGES{here denoted by $SS_{\mathrm{res}}(\cdot)$,} as given in Eq.~\eqref{eq:partialF}:
{\small
\begin{equation}
\label{eq:partialF}
F \;=\; \frac{\big(SS_{\mathrm{res}}(\mathcal{M}_0) - SS_{\mathrm{res}}(\mathcal{M}_1)\big) / (V_1 - V_0)}
{SS_{\mathrm{res}}(\mathcal{M}_1) / (N - V_1 - 1)} ,
\end{equation}
}
\CHANGES{where $i\in\{0,1\}$ indexes the reduced and full models, $N$ is the sample size, and $V_i$ is the number of predictors in $\mathcal{M}_i$ (excluding the intercept). Under $\mathcal{M}_0$ and standard regularity assumptions, $F$ follows an $F$ distribution with numerator and denominator degrees of freedom (df) $(V_1-V_0,\;N-V_1-1)$.} A large $F$ indicates that the added terms (e.g., humidity or an interaction such as distance$\times$humidity) reduce unexplained variance beyond chance.

Since indoor datasets are often unbalanced and predictors can be correlated (e.g., temperature with humidity), the choice of ANOVA type matters \cite{langsrudANOVAUnbalancedData2003}. \textit{Type~I} (sequential) ANOVA attributes sums of squares in the order variables enter the model and is order-dependent. It is mainly used for designed and balanced experiments. \textit{Type~II} ANOVA evaluates each main effect after adjusting for the other main effects (but not interactions) and is preferred when interactions are absent or excluded. \textit{Type~III} ANOVA tests each effect while adjusting for all other main effects and interactions, and is common when interaction terms are included. Across types, collinearity inflates uncertainty and can obscure effects. For this reason, studies often report variance inflation factors (VIFs), standardized predictors, and interpret \textit{Type}~II and \textit{Type}~III results with caution.

Inference assumes residuals are approximately Gaussian, variance-stable, and weakly autocorrelated. When diagnostics suggest deviations, several robustifications appear in the literature. Heteroskedasticity-consistent (HC) covariance estimators for $t$- and $F$-tests, permutation (randomization) ANOVA in small or non-normal samples, and block-bootstrap CIs when short-memory temporal dependence is present \cite{langsrudANOVAUnbalancedData2003}. Nonparametric rank tests (e.g., Kruskal--Wallis) are used for group-wise comparisons when normality is doubtful, complementing rather than replacing parametric ANOVA.

Empirical indoor work illustrates these points through varied designs. Two-way ANOVA has been used to evaluate the effects of categorized temperature and humidity on RSSI in mote-based testbeds, finding significant main effects \cite{chellougImpactTemperatureHumidity2014}. Full-factorial analyses have quantified interactions among channel, link path, and transmit power, with interaction terms explaining sizable portions of variability \cite{christmannExperimentalDesignAnalysis2010}. ANOVA has also appeared in network-level studies, for example, to identify redundant sensing clusters, demonstrating statistically grounded routes to energy savings \cite{harbEnhancedKMeansANOVABased2015}. Collectively, these practices establish a procedure that begins with a physically motivated linear specification, tests main effects with \textit{Type}~II or \textit{Type}~III ANOVA as design dictates, inspects residual assumptions, and deploys robust or rank-based alternatives when diagnostics warrant.

\subsection{Residual Distribution Diagnostics}
\label{subsec:shadow}

Residual distribution (shadow fading) refers to the residual variability that remains after accounting for distance, structural losses, and auxiliary predictors, such as the indoor microclimate \cite{gonzalez-palacioMachineLearningBasedCombinedPath2023}. In indoor settings, these residuals often deviate from the Gaussian assumption. Specifically, skewness and heavy tails arise due to dynamic blockage and hardware heterogeneity, while multimodality reflects regime changes, such as occupancy cycles or HVAC operation. A defensible characterization, therefore, considers distributional shape, variance stability, and potential temporal or group dependence using complementary parametric and nonparametric tools, as summarized in Table~\ref{tab:residual_methods}. In practice, to avoid optimistic bias, all diagnostic fits and tail summaries are computed on OOF residuals.

\begin{table}[hbt!]
\centering
\caption[Residual diagnostic methods used in indoor path loss modeling.]{\textbf{Residual diagnostic methods used in indoor path loss modeling.} In the Type column, a circle (\protect\tikz \protect\draw (0,0) circle (.11cm);) marks parametric methods, whereas a square (\protect\tikz \protect\draw (0,0) rectangle (.21cm,.21cm);) marks nonparametric methods; i.i.d.\ means independent and identically distributed.}
\label{tab:residual_methods}
\begin{tabularx}{\linewidth}{
  >{\centering\arraybackslash}p{3.3cm}
  >{\centering\arraybackslash}p{0.7cm}
  >{\raggedright\arraybackslash}X
}
\toprule
\textbf{Method} & \textbf{Type} & \textbf{Purpose and typical use} \\
\midrule
Single-family goodness-of-fit  & \tikz\draw (0,0) circle (.11cm); &
Screen for Gaussianity versus heavy tails/asymmetry using Q--Q plots plus omnibus tests; establishes whether a unimodal law suffices \cite{allenIndoortooutdoorEmpiricalPath2017} \\
\midrule
Gaussian Mixture Models (GMMs) & \tikz\draw (0,0) circle (.11cm); &
Compact representation of heterogeneous regimes (e.g., occupancy/HVAC states); select components by Akaike information criterion (AIC) or Bayesian information criterion (BIC) and validate on held-out data (e.g., likelihood) \cite{wangGreyModelMixture2020} \\
\midrule
Kernel Density Estimation (KDE) & \tikz\draw (0,0) rectangle (.21cm,.21cm); &
Distribution visualization without parametric constraints; reveals secondary modes or mild asymmetry that guide model choice \cite{silvermanDensityEstimationStatistics2018} \\
\midrule
Heteroskedasticity tests & \tikz\draw (0,0) circle (.11cm); &
Detect variance changes with fitted values or covariates; motivates weighting or robust standard errors when variance is not constant \cite{astiviaHeteroskedasticityMultipleRegression2019} \\
\midrule
\CHANGES{Group-wise location/dispersion tests} & \tikz\draw (0,0) rectangle (.21cm,.21cm); &
\CHANGES{Kruskal–Wallis for median/location differences; Brown–Forsythe (median‑centered Levene) or Fligner–Killeen for dispersion.} \cite{chanLearningUnderstandingKruskalWallis1997} \\
\midrule
Autocorrelation diagnostics  & \tikz\draw (0,0) circle (.11cm); &
Assess short-memory dependence that affects uncertainty estimation and motivates blocked validation protocols \cite{zhangTimeSeriesBasedStudy2021} \\
\midrule
Bootstrap quantiles (i.i.d.\ and block) & \tikz\draw (0,0) rectangle (.21cm,.21cm); &
Empirical confidence intervals (CIs) for tail quantiles used in fade margin budgeting; block variants preserve dependence \cite{efronBootstrapMethodsAnother1979} \\
\bottomrule
\end{tabularx}
\end{table}

\begin{table}[htbp]
\centering
\caption[A comparative summary of existing indoor LoRaWAN propagation studies.]{\textbf{A comparative summary of existing indoor LoRaWAN propagation studies.} We compare prior work by assessing whether environmental factors (EF), advanced modeling (AM), and residual analysis (RA) were addressed. For \textbf{EF}, \protect\ding{51} indicates included, I indicates indirect treatment, and \protect\ding{55} indicates not included; for \textbf{AM}, \protect\ding{51} indicates performed, P indicates partially performed, and \protect\ding{55} indicates not performed; for \textbf{RA}, \protect\ding{51} indicates performed and \protect\ding{55} indicates not performed.}

\label{tab:literature_comparison}
\begin{tabularx}{\linewidth}{
  >{\centering\arraybackslash}m{0.05\textwidth}
  >{\raggedright\arraybackslash}m{0.75\textwidth}
  >{\centering\arraybackslash}m{0.02\textwidth}
  >{\centering\arraybackslash}m{0.02\textwidth}
  >{\centering\arraybackslash}m{0.02\textwidth}
}
\toprule
\textbf{Ref.} & \textbf{Focus and summary of findings} & \textbf{EF} & \textbf{AM} & \textbf{RA} \\
\midrule
\cite{ayelePerformanceAnalysisLoRa2017} (2017) &
Single‑floor office tests ($868\,\mathrm{MHz}$) measuring RSSI and packet delivery vs. spreading factor (SF) 7–12 across four locations; higher SF did not uniformly improve delivery (long time-on-air effects), offering practical configuration guidance &
I & \ding{55} & \ding{55} \\
\midrule
\cite{hosseinzadehNeuralNetworkPropagation2017} (2017) &
Outdoor–indoor office study: adjusted COST 231 MWM with a neural network residual corrector; test error improves (mean squared error from $21.0$ to $11.23\,\mathrm{dBm^2}$) when generalizing from the 8th to the 7th floor &
\ding{55} & \ding{51} & \ding{55} \\
\midrule 
\cite{petajajarviEvaluationLoRaLPWAN2017} (2017) &
Campus‑scale indoor trials ($868\,\mathrm{MHz}$) with a Kerlink gateway ($\approx 24\,\mathrm{m}$) and static/wearable nodes; reliable room‑level coverage at SF12, $14\,\mathrm{dBm}$ with packet delivery $\approx 96.7\%$; descriptive path‑loss statistics &
\ding{55} & \ding{55} & \ding{55} \\
\midrule
\cite{erbatiAnalysisLoRaWANTechnology2018} (2018) &
Indoor multi‑floor measurements (Duisburg): same‑floor RSSI/SNR largely SF‑independent; basement delivery ranged $62\%$ (SF7) to $100\%$ (SF10/SF12); larger payloads reduced delivery &
\ding{55} & \ding{55} & \ding{55} \\
\midrule
\cite{elchallLoRaWANNetworkRadio2019} (2019) &
Indoor empirical path loss modeling at $868\,\mathrm{MHz}$: COST 231 MWM and floor with $n=2.85$, $PL_{0}=120.4\,\mathrm{dB}$, per-wall $L_{w}=1.41\,\mathrm{dB}$, per-floor $L_{f}=10\,\mathrm{dB}$, and $b=0.47$; reports shadowing $\sigma\approx 8\text{–}9.7\,\mathrm{dB}$ &
\ding{55} & P & \ding{55} \\
\midrule
\cite{muzammirPerformanceAnalysisLoRaWAN2019} (2019) &
Indoor multi-floor study (gateway on Level 9; nodes on Levels 5–9): measured RSSI and packet delivery vs.\ SF7–SF12 ($125\,\mathrm{kHz}$) and payload; delivery drops with floor separation and larger payloads, time-on-air rises at high SF &
\ding{55} & \ding{55} & \ding{55} \\
\midrule
\cite{muppalaInvestigationIndoorLoRaWAN2021} (2021) &
Three‑storey office mapping ($N=89$ points): hallway RSSI–distance shows anomalies at $\sim\!40,80,120\,\mathrm{m}$ from window reflections; brick walls cause strong short‑range loss &
\ding{55} & \ding{55} & \ding{55} \\
\midrule
\cite{sabanExperimentalAnalysisIoT2021a} (2021) &
Indoor multi‑block building (SF12, $14\,\mathrm{dBm}$): near‑perfect delivery within the gateway block; first losses at $\sim 40\,\mathrm{m}$, steep drop beyond $\sim 75\,\mathrm{m}$; recommends $\lesssim 70\,\mathrm{m}$ links &
\ding{55} & \ding{55} & \ding{55} \\
\midrule
\cite{harindaPerformanceLiveMultiGateway2022} (2022) &
Live indoor multi-gateway office (EU-868, 26\,days): first-attempt success $99.95\%$ indoors vs $95.7\%$ outdoors; city-site scans show interference up to $97.3\%$ (uplink) and $54\%$ (downlink) &
\ding{55} & \ding{55} & \ding{55} \\
\midrule
\cite{robles-encisoLoRaZigbee5G2023a} (2023) &
Indoor corridor at $868\,\mathrm{MHz}$: fits the close-in reference and floating-intercept forms to LoRa (plus Zigbee/5 G) in two $\mathrm{NLoS}$ zones; finds $n<2$ (waveguide-like) in $\mathrm{NLoS}\text{-}1$ and $n>3$ in $\mathrm{NLoS}\text{-}2$; floating-intercept forms outperforms close-in reference in $\mathrm{NLoS}\text{-}2$ and is used to relate received power to bit error rate thresholds &
\ding{55} & P & \ding{55} \\
\midrule
\cite{alkhazmiAnalysisRealWorldLoRaWAN2023} (2023) &
Indoor 8-story building with rooftop end device, gateway moved floor-by-floor; fixed SF7, $14\,\mathrm{dBm}$, $125\,\mathrm{kHz}$, $868\,\mathrm{MHz}$; 50 packets/floor at 5-s intervals; RSSI/SNR rise with proximity (e.g., Floor 1 $\mathrm{RSSI}\approx-110\,\mathrm{dBm}$, and Floor 8 $\approx-71\,\mathrm{dBm}$); near-zero loss on Floors 2–8 &
\ding{55} & \ding{55} & \ding{55} \\
\midrule
\cite{aksoyComparativeAnalysisEnd2024} (2024) &
For an 18-floor indoor building with a rooftop gateway, compares end device logs (1 day) vs Adeunis Field Test Device (FTD) (100 samples/floor) at Floors 1/6/12/18; reports RSSI/SNR trends with distance and SF allocations; links operate at negative SNR down to $-13.8\,\mathrm{dB}$ &
\ding{55} & \ding{55} & \ding{55} \\
\midrule
\cite{voAdvancePathLoss2024} (2024) &
For an indoor lab: proposes a dynamic log-distance model with an additive noise term $T$ and Kalman filtering; estimates indoor path loss exponent $n=2.103$; mean distance error $0.565\,\mathrm{m}$ with $90\%$ of errors $<1.08\,\mathrm{m}$ over $3\text{–}7\,\mathrm{m}$ &
\ding{55} & \ding{51} & \ding{55} \\
\midrule
\cite{mokuaobiriComprehensiveDataDescription2025} (2025) &
\CHANGES{Our dataset descriptor for the deployment (indoor office, 8th floor), described in this work. We compare a structure-only multi-wall log-distance baseline against an environment-augmented variant (environmental sensing and SNR), improving $5$-fold cross-validation RMSE from $10.577$ to $8.034\,\mathrm{dB}$ and $R^2$ from $0.691$ to $0.822$} &
\ding{51} & P & \ding{55} \\
\midrule
\cite{obiriStatisticalEvaluationIndoor2025a} (2025) &
\CHANGES{We perform early statistical evaluation on the campaign in \cite{mokuaobiriComprehensiveDataDescription2025} using MLR with ANOVA, and residual distribution analysis. Adding environment-aware covariates materially reduced unexplained variance and motivated multimodal shadowing models beyond Gaussian residual assumptions} &
\ding{51} & P & \ding{51} \\
\midrule
\emph{This Work} &
Using a $12$-month dataset with multivariate path loss with distance, walls/floors, and environment variables, we (i) compare MLR vs.\ second-order polynomial and regularized/Bayesian variants via cross-validation, (ii) perform residual diagnostics (GMM, KDE/bootstrapping, Kruskal–Wallis), and (iii) pursue quantile-based fade margin calibration on OOF residuals &
\ding{51} & \ding{51} & \ding{51} \\
\bottomrule
\end{tabularx}
\end{table}

Regarding distributional shape, studies fit unimodal families such as the Normal, Student’s $t$, and Skew--Normal (asymmetry) \cite{azzaliniSkewNormalRelatedFamilies2014}, and occasionally Cauchy (very heavy tails) \cite{johnsonContinuousUnivariateDistributions1995}, and inspect Q--Q plots alongside omnibus tests (e.g., Kolmogorov--Smirnov (KS) \cite{masseyjr.KolmogorovSmirnovTestGoodness1951}) to gauge tail thickness and asymmetry. When a single law is inadequate, GMMs are used to represent heterogeneous regimes, with the component count selected by AIC or BIC and validated on held-out data \cite{reynoldsGaussianMixtureModels2009}. Nonparametric KDE often reveals subsidiary modes or mild asymmetry that inform model choice \cite{silvermanDensityEstimationStatistics2018}. Variance heterogeneity is probed with scale–location plots and classical tests (Breusch--Pagan/White). Grouped comparisons by device, location, or time-of-day use Levene/Brown--Forsythe or rank-based tests (Kruskal--Wallis) to detect structure-induced dispersion shifts \cite{chanLearningUnderstandingKruskalWallis1997}. Temporal diagnostics such as the autocorrelation function (ACF) and partial autocorrelation function (PACF), the Durbin–Watson statistic, and the Ljung–Box test check short-memory effects that can bias uncertainty estimates if ignored \cite{zhangTimeSeriesBasedStudy2021}. For uncertainty in tail summaries, bootstrap resampling provides empirical CIs for error quantiles \cite{efronBootstrapMethodsAnother1979}, and block variants preserve short-range dependence when it is present.

In practice, the aim is not maximal flexibility but the simplest residual model consistent with diagnostics and calibrated tails. Indoors, this often means starting from a Normal fit, escalating to $t$ or Skew--Normal when tails or asymmetry dominate, and adopting a low-order GMM only when multimodality is persistent and interpretable. The resulting upper-tail quantiles directly support fade margin budgeting in reliability analyses, provided they are derived from OOF residuals.

\subsection{Design Goals and Methodological Positioning}
\label{subsec:design_goals}

Across prior work (see Table~\ref{tab:literature_comparison}), persistent gaps include the limited integration of environmental context, scarce leakage-safe evaluation with calibrated uncertainty, and residual shapes that are seldom scrutinized beyond simple normality checks. Consequently, fade margin prescriptions are rarely validated on held-out data. Our study addresses these gaps with a physics-grounded mean, selective nonlinear terms, time‑blocked cross‑validation, and residual diagnostics to obtain interpretable effects and calibrated reliability margins. Guided by these gaps, our design goals are to: \textbf{(i)} preserve physical interpretability so that distance, walls, and environmental contributions remain attributable, \textbf{(ii)} incorporate environmental context to reduce unexplained variance rather than absorbing it into shadowing, \textbf{(iii)} enforce leakage-safe evaluation so reported gains reflect true generalization, and \textbf{(iv)} quantify predictive uncertainty for reliability budgeting, not just point accuracy.

These goals are operationalized with a mean specification anchored in Eq.~\eqref{eq:pl-mw-en} and a selective quadratic extension in Eq.~\eqref{eq:pl-mw-en-quad} applied to the continuous predictors (distance, environmental factors, and the SNR, while wall counts and weights remain additive and linear. Polynomial expansion precedes standardization, while all preprocessing (feature mapping and scaling) is confined within cross-validation folds (time-blocked) to prevent leakage, and diagnostics and fade margin quantiles are computed on OOF residuals. Three model families are compared: a linear baseline MLR, the second-order polynomial design, and a BLR model (conjugate NIG and Zellner $g$-prior) on the linear design for calibrated predictive uncertainty. Penalized variants (Ridge, Lasso, and Elastic Net) are tuned by cross-validation with parsimony-oriented selection.

\begin{figure}[hbt!]
    \centering
    \includegraphics[width=\linewidth]{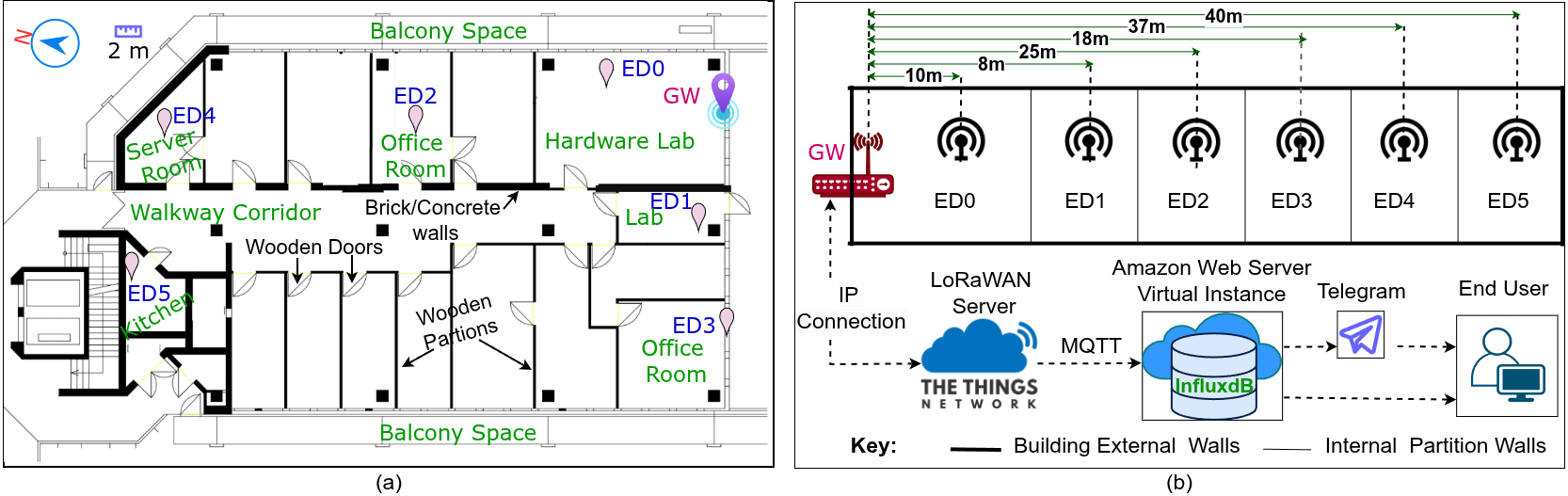}
    \caption{\CHANGES{\textbf{Overview of the indoor LoRaWAN measurement campaign:} (a) Floor plan with the gateway (GW) and six end devices (ED0–ED5), annotated obstructions distinguish brick/reinforced-concrete walls and wooden partitions/doors. Scale bar: $2\,\mathrm{m}$; icons not to scale. (b) Data acquisition chain, forwarding uplinks from The Things Network (TTN) to an InfluxDB time-series database via Message Queuing Telemetry Transport (MQTT).}}

    \label{fig:floor_plan}
\end{figure}

\section{Methodology}
\label{sec:methodology}

\subsection{Experimental Setup}
\label{subsec:exp_setup}
In this section, we describe the deployment geometry, sensing hardware, radio configuration, and the backend pipeline used to acquire the dataset for our path loss modeling and analysis.

\subsubsection{Site and Layout}
\CHANGES{This data measurement campaign builds on the indoor LoRaWAN deployment reported in our EU $868\,\mathrm{MHz}$ dataset descriptor \cite{mokuaobiriComprehensiveDataDescription2025}. We summarize the elements required to reproduce the measurement conditions and the end-to-end data logging pipeline. Measurements were collected on the 8th floor of an academic building ($240\,\mathrm{m}^2$ and approx.\ $250\,\mathrm{m}$ above sea level). The floor follows a corridor-centric office layout with a semi-open central walkway (Fig.~\ref{fig:floor_plan}(a)). Structural walls are brick and reinforced concrete, while internal separations are primarily lightweight wooden partitions and wooden doors. This mixture yields links spanning LoS and NLoS, with ED0 being the only LoS device to the gateway. End device to gateway distances span $d \in [8,40]\,\mathrm{m}$, and each link is tagged in Fig.~\ref{fig:floor_plan}(a) with its wall-count tuple $\mathbf{w}=(W_{\mathrm{brick}},W_{\mathrm{wood}})$: ED0 $(0,0)$, ED1 $(1,0)$, ED2 $(0,2)$, ED3 $(1,2)$, ED4 $(0,5)$, ED5 $(2,2)$.}

\CHANGES{Measurements were collected during normal use of offices, labaratories, and shared spaces (uncontrolled human presence and mobility), spanning both working-hour and off-hour regimes with intermittent corridor traffic. Human presence and movement were not scripted, and no fixed headcount is assumed. We therefore treat occupancy as an uncontrolled driver of indoor non-stationarity and capture its effect indirectly through the co-recorded environmental time series. In practice, indoor CO$_2$ exhibits strong diurnal structure with peaks that plausibly track human activity. At the same time, PM\textsubscript{2.5} and other measurements can also be shaped by operational factors such as heating, time of day, and window/door opening \cite{grubelDenseIndoorSensor2022b}. Since headcount was not instrumented, these variables serve as proxy signals for activity-driven dynamics that may co-vary with LoRaWAN link reliability in situ.}

\begin{figure}[hbt!]
    \centering
    \includegraphics[width=\linewidth]{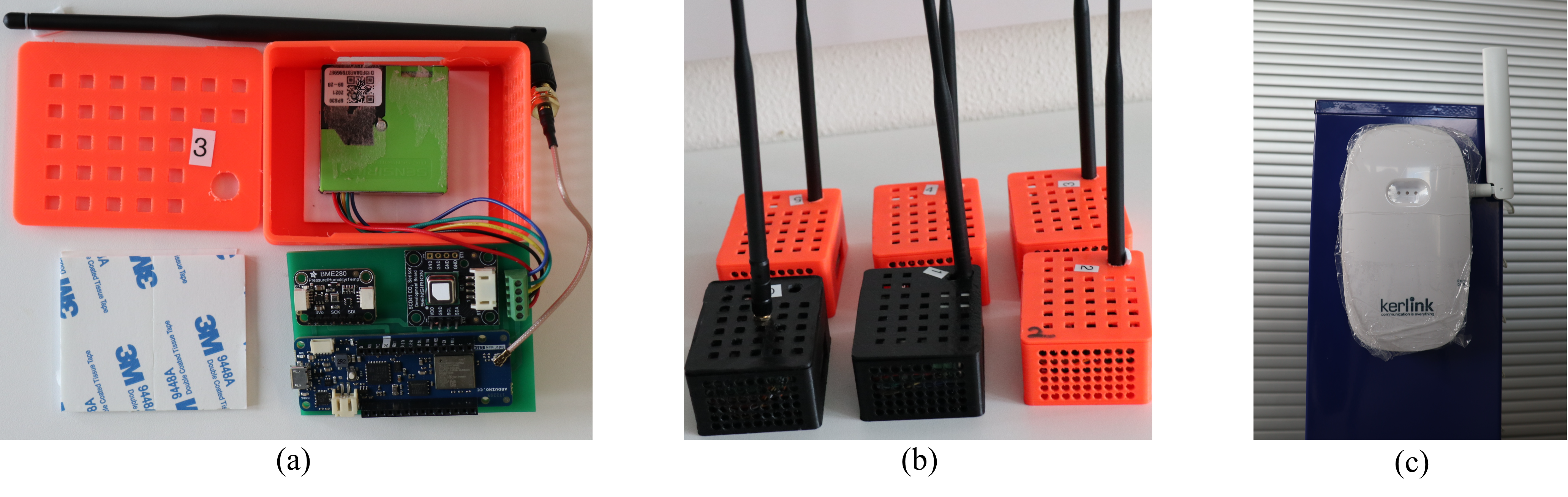}
    \caption{\CHANGES{\textbf{LoRaWAN end devices and indoor gateway used in the deployment.} (a) End device internals: Arduino MKR WAN 1310 (Murata LoRa, EU $868\, \mathrm{MHz}$) with BME280 (pressure), SCD41 (CO\textsubscript{2}, temperature, humidity), and SPS30 (PM\textsubscript{2.5}). (b) Assembled end devices in 3D-printed enclosures. (c) Kerlink Wirnet iFemtoCell indoor gateway (wall-mounted).}}
    \label{fig:end_device}
\end{figure}

\subsubsection{End Devices}
\CHANGES{We deployed six identical end devices (EDs, ED0--ED5) at fixed locations for the duration of the campaign (Fig.~\ref{fig:floor_plan}(a)). Each end device was built around an Arduino MKR WAN 1310 (Murata LoRa, EU $868\,\mathrm{MHz}$) and housed in a ventilated 3D-printed enclosure (Fig.~\ref{fig:end_device}(a) and (b)). The platform exposes standard peripheral buses, namely Inter-Integrated Circuit (I2C), Serial Peripheral Interface (SPI), and Universal Asynchronous Receiver/Transmitter (UART), which simplifies clean sensor integration and debugging without additional interface hardware \cite{gonzalez-palacioLoRaWANPathLoss2023}. In this deployment, all sensors (Table~\ref{tab:sensors}) were connected to a shared I2C bus to minimize wiring and ensure consistent sampling. For radio transmission, each ED uses an external omnidirectional antenna (peak gain $G_{\mathrm{tx}} \approx 0.4\,\mathrm{dBi}$) connected via a short pigtail ($L_{tx} \approx 0.14 \,\mathrm{dB}$) and transmits at $P_{tx}\approx 14\,\mathrm{dBm}$. Antennas were mounted vertically at approximately $0.8\,\mathrm{m}$ above floor level to reflect typical indoor IoT installations.}

\begin{table}[hbt!]
\centering
\caption{\CHANGES{\textbf{Sensor suite integrated into each end device.} The table summarizes the sensing modules, their operating interfaces or measurement principles, and the variables captured with their reported sensitivity. I2C refers to Inter-Integrated Circuit, SPI to Serial Peripheral Interface, and UART to Universal Asynchronous Receiver/Transmitter.}}
\label{tab:sensors}
\begin{tabularx}{\linewidth}{
  >{\centering\arraybackslash}p{0.15\linewidth}
  >{\raggedright\arraybackslash}p{0.28\linewidth}
  >{\raggedright\arraybackslash}X
}
\toprule
\textbf{Sensor} & \textbf{Interface / principle} & \textbf{Measured variables and sensitivity} \\
\midrule
Sensirion SCD41
& I2C (used); UART is supported; includes photoacoustic CO\textsubscript{2} sensing with integrated temperature and relative humidity channels; automatic self-calibration
& Measures, \textbf{(i)} carbon dioxide concentration: $\SIrange{400}{5000}{ppm}$; accuracy $\pm(\SI{40}{ppm} + 5\%~\text{of reading})$, \textbf{(ii)} temperature: $\SIrange{-10}{60}{\degreeCelsius}$; accuracy $\pm\SI{1.5}{\degreeCelsius}$ (typ.\ $\pm\SI{0.8}{\degreeCelsius}$ for $\SIrange{15}{35}{\degreeCelsius}$), and \textbf{(iii)} relative humidity: $\SIrange{0}{100}{\percent\relax\mathrm{RH}}$; accuracy $\pm\SI{9}{\percent\relax\mathrm{RH}}$ (typ.\ $\pm\SI{6}{\percent\relax\mathrm{RH}}$ for $\SIrange{15}{35}{\degreeCelsius}$ and $\SIrange{20}{65}{\percent\relax\mathrm{RH}}$)\\
\midrule
Adafruit BME280
& I2C (used); SPI supported; pressure sensing based on Bosch technology
& Measures barometric pressure in the range of $300$--$1100$\,hPa; absolute accuracy $\pm 1.0$\,hPa \\
\midrule
Sensirion SPS30
& I2C (used); UART supported; laser-scattering particulate sensing with contamination-resistance; self cleaning fan
& Measures particulate matter mass concentration (PM\textsubscript{2.5}) in the range of $0$--$1000\,\mu$g/m$^3$; its reported accuracy is $\pm 10$\% for the range within $100$--$1000\,\mu$g/m$^3$ size of particles \\
\bottomrule
\end{tabularx}
\end{table}

\subsubsection{Gateway and Backhaul}
\CHANGES{We used a Kerlink Wirnet iFemtoCell indoor gateway (GW, Fig.~\ref{fig:end_device}(c)), wall-mounted at $1.0\,\mathrm{m}$ above floor level. The GW was equipped with a vertically polarized omnidirectional rubber-dipole antenna (peak gain $G_{\mathrm{rx}} \approx 3\,\mathrm{dBi}$) connected directly at the RF port, hence receiver-side cable losses $L_{\mathrm{rx}}$ were negligible. In its EU868 configuration, the gateway provides a receiver sensitivity ($S$) up to $-141\,\mathrm{dBm}$ at SF12 ($125\,\mathrm{kHz}$ bandwidth (BW)), enabling robust packet capture on heavily obstructed links. As was deployed in \cite{obiriLongRangeWideArea2023}, uplinks were forwarded over an Ethernet backhaul to The Things Network (TTN) stack (v3.35.2). The end devices operated under the LoRaWAN MAC specification v1.0.2, and TTN exposed per-packet metadata required for link analysis (timestamps, SF, frequency, RSSI, and SNR).}

\subsubsection{Radio Configuration and Measurement Loop}
\CHANGES{Each end device followed a deterministic one-minute cycle: \textbf{(i)} it sampled the sensor suite, \textbf{(ii)} packed the readings into a compact 18-byte binary frame, and \textbf{(iii)} transmitted a single uplink. The frame carried temperature (14\, bit), relative humidity (14\, bit), barometric pressure (17\, bit), carbon dioxide concentration (13\, bit), particulate matter (PM\textsubscript{2.5}, 17\, bit), and a monotonic packet counter (32\, bit), totaling 107\, bit ($\approx$14\, bytes), with the remaining 37\, bit reserved as padding and for future extensions. Uplinks were sent at a transmit power of $14\,\mathrm{dBm}$ in the EU868 band. We cycled the spreading factor (SF) from SF7 to SF10 to probe a practical robustness–airtime trade-off, since higher SF increases robustness at the cost of longer transmissions \cite{grubelDenseIndoorSensor2022b}. With a $125\,\mathrm{kHz}$ bandwidth (coding rate set to 4/5) and an 18-byte payload, using the standard LoRa time-on-air model with an header, cyclic redundancy check (CRC) enabled, and an 8-symbol preamble, the time-on-air remains bounded (approximately $51\,\mathrm{ms}$ at SF7 and $330\,\mathrm{ms}$ at SF10). This supports the $60\,\mathrm{s}$ reporting interval under EU868 duty-cycle constraints \cite{chaudhariLPWANTechnologiesIoT}, and reduces self-induced congestion.}

\subsubsection{Data Pipeline and Storage}
\CHANGES{As shown in Fig.~\ref{fig:floor_plan}(b) (adopted from \cite{obiriLongRangeWideArea2023, mokuaobiriComprehensiveDataDescription2025}), uplinks arriving at the TTN server were decoded using an application-level JavaScript payload formatter and streamed via Message Queuing Telemetry Transport (MQTT) (v3.1.1). A subscriber running on an Amazon Web Services Elastic Compute Cloud (AWS EC2) instance ingested the MQTT stream and persisted both decoded sensor fields and TTN metadata into an InfluxDB (v1.8.9) time-series database. The pipeline operated continuously from October 2024 to September 2025, yielding a dense, time-aligned record of environmental conditions and link state for subsequent modeling and analysis.}

\subsection{Path Loss Modeling}
\label{subsec:model_fitting}

\CHANGES{Path loss modeling in LoRaWAN is a link-budget exercise, because meeting a target reliability means the transmit setting (e.g., SF) must cover the mean propagation loss plus a safety headroom. We index the end device to the gateway link by $\ell$ and the received packets by $i$. Let $L_{\ell}(d,f,\ldots)$ denote the (random) path loss on link $\ell$ under propagation conditions $(d,f,\ldots)$.  In link-budget form, successful reception requires that Eq.~\eqref{eq:lb_planning} holds:}
\begin{equation}
\label{eq:lb_planning}
P_{\mathrm{tx}}\;\ge\;L_{\mathrm{tx}}-G_{\mathrm{tx}} +L_{\ell}(d,f,\ldots)-G_{\mathrm{rx}}+L_{\mathrm{rx}}
+M_{\mathrm{L}}+S(\mathrm{SF},\mathrm{BW}),
\end{equation}
\CHANGES{where $P_{\mathrm{tx}}$ is the transmit power, $G_{\mathrm{tx}}$ and $G_{\mathrm{rx}}$
are antenna gains, $L_{\mathrm{tx}}$ and $L_{\mathrm{rx}}$ are feeder/connector losses,
$S(\mathrm{SF},\mathrm{BW})$ is the receiver sensitivity, and $M_{\mathrm{L}}$ is the total link margin. \cite{gonzalez-palacioMachineLearningBasedCombinedPath2023}. All power units are in $\mathrm{dBm}$, while all losses and margins are in $\mathrm{dB}$. In LoRaWAN's adaptive data rate (ADR), this headroom is implemented via the SNR-margin rule, with a fixed fade margin constant (often set to $10\,\mathrm{dB}$). We keep the same principle but replace the fixed hedge by a data-driven fade margin inferred from the upper tail of the OOF residuals after fitting $\widehat{L}(\cdot)$ (formalized in Sec.~\ref{subsec:fm}). Note that in this work, we interpret the link margin as $M_{\mathrm{L}}=M_{\mathrm{F}}+M_{\mathrm{I}}$, where $M_{\mathrm{F}}$ is the calibrated fade margin and $M_{\mathrm{I}}$ collects any additional implementation or interference margins. In our evaluation, we set $M_{\mathrm{I}}=0$ to isolate the propagation-driven component. To anchor the analysis to our dataset, we construct the experimental path loss used as the regression response directly from TTN metadata by treating the gateway-reported RSSI as a received-power estimate. Let $P_{{\mathrm{rx}},i}$ denote the gateway-reported received power for packet $i$ (reported as RSSI in TTN metadata). Equation~\eqref {eq:pl_from_rssi} gives the per-packet path loss:}
\begin{equation}
\label{eq:pl_from_rssi}
L_{\ell,i}= P_{\mathrm{tx}}-L_{\mathrm{tx}}+G_{\mathrm{tx}}+G_{\mathrm{rx}}-L_{\mathrm{rx}}-P_{{\mathrm{rx}},i}.
\end{equation}
\CHANGES{While COST 231 MWM (log-distance) \cite{europeancommissionCOSTAction2311999} captures geometry and obstruction losses, it does not represent the occupied-building dynamics that drive indoor non-stationarity, so leaving these effects in the residual inflates the margin required for high reliability.} 
We therefore augment it with an environmental state vector and SNR as a link-state indicator to capture indoor non-stationarity.  Let $\widehat{L}_{\ell,i}$ denote a model-based mean prediction for $L_{\ell,i}$  for packet $i$ at distance $d$ (m) relative to a reference distance $d_0=1\, \mathrm{m}$, and let $f$ denote the carrier frequency (MHz). \CHANGES{Let $W_k$ be the number of penetrated obstructions of wall type $k\in\{1,\ldots,K\}$ (with loss $\omega_k$ in $\mathrm{dB}$ per wall), so that $\mathbf{w}=(W_1,\ldots,W_K)^\top$ (in our site, $\mathbf{w}=(W_{\mathrm{brick}},W_{\mathrm{wood}})^\top$). Let $E_j$ be the $j$-th environmental covariate ($j\in\{1,\ldots,P\}$), stacked as $\mathbf{e}=(E_1,\ldots,E_P)^\top$, and let $\gamma$ denote the gateway-reported SNR (in $\mathrm{dB}$).} The additive linear mean model is given in Eq.~\eqref{eq:pl-mw-en} as,
{\small
\begin{equation}
\label{eq:pl-mw-en}
L_{\ell,i}
=
\underbrace{
\beta_0
+ 10n\log_{10}\!\left(\frac{d_i}{d_0}\right)
+ 20\log_{10}(f_i)
+ \sum_{k=1}^{K} \omega_k\,W_{k,i}
+ \sum_{j=1}^{P} \varepsilon_j\,E_{j,i}
+ k_{\gamma}\,\gamma_i
}_{\widehat{L}_{\ell,i}}
+ \psi_i ,
\end{equation}
}
where $\beta_0$ is the intercept and $n$ is the path loss exponent, while the term $20\log_{10}(f)$ accounts for frequency-dependent free-space loss \cite{friisNoteSimpleTransmission1946}. In our single-band EU868 deployment, it is constant and is absorbed into $\beta_0$ during fitting. \CHANGES{The coefficient $\varepsilon_j$ weights the environmental covariates $E_j$ (for readability in later tables, we use the following mnemonic subscripts for the coefficients: $\varepsilon_{\mathrm{C}}$ for CO$_2$, $\varepsilon_{\mathrm{RH}}$ for relative humidity, $\varepsilon_{\mathrm{PM}}$ for PM\textsubscript{2.5}, $\varepsilon_{\mathrm{BP}}$ for barometric pressure, and $\varepsilon_{\mathrm{T}}$ for temperature).} The factor $k_{\gamma}$ scales the link-state term $\gamma$, and $\psi$ denotes shadowing.  We include $\gamma$ as a practical receiver-side indicator that captures instantaneous demodulation margin and channel interference conditions \cite{szafranskiPredictabilityLoRaWANLink2024}. In outdoor LoRaWAN campaigns, $\gamma$ (SNR) is likewise used alongside RSSI to derive path loss from gateway metadata \cite{biancoLoRaSystemSearch2021}. More generally, poorer $\gamma$ aligns with higher path loss, i.e., a negative SNR effect \cite{gonzalez-palacioLoRaWANPathLoss2023}. When $\gamma$ is unavailable for pure planning, the same pipeline applies after dropping this term, with fade margins increasing modestly because the mean absorbs less variability.

To allow mild curvature and context coupling while preserving the additive multi-wall structure, we extend Eq.~\eqref{eq:pl-mw-en} with a restricted second-order polynomial applied only to the continuous predictors. Define $z_d \equiv 10\log_{10}(d/d_0)$ and $\mathbf{u}\equiv[z_d,\ \mathbf{e}^\top,\ \gamma]^\top \in \mathbb{R}^{q}$, where $q=P+2$. Walls remain purely additive (no wall squares and no wall interactions). Under this restriction, the polynomial model (Eq.~\eqref{eq:pl-mw-en-quad}) is written as:
{\small
\begin{equation}
\label{eq:pl-mw-en-quad}
L_{\ell,i}=\underbrace{\widehat{L}_{\ell,i}+\Delta_{\mathrm{poly}}(\mathbf{u}_i)}_{\widehat{L}_{\ell,\mathrm{poly},i}}+\psi_i,\qquad
\Delta_{\mathrm{poly}}(\mathbf{u}_i)=\sum_{a=1}^{q}\tilde{h}_{aa}\,u_{a,i}^2+\sum_{a=1}^{q-1}\sum_{b=a+1}^{q}\tilde{h}_{ab}\,u_{a,i}u_{b,i},
\end{equation}
}
\CHANGES{where $\{\tilde{h}_{ab}\}_{1\le a\le b\le q}$ are unique quadratic coefficients in one-to-one correspondence with the monomials $\{u_a^2\}$ and $\{u_a u_b\}_{a<b}$ among the continuous predictors, and where $u_a$ denotes the $a$-th component of $\mathbf{u}$ with indices $a,b\in\{1,\ldots,q\}$. Equivalently, $\Delta_{\mathrm{poly}}(\mathbf{u})=\mathbf{u}^\top H\,\mathbf{u}$ with $H=H^\top$ (a symmetric coefficient matrix), where $h_{aa}=\tilde{h}_{aa}$ and $h_{ab}=\tilde{h}_{ab}/2$ for $a<b$. This introduces quadratic terms among the continuous predictors (e.g., $z_d^2$, $E_j^2$, $\gamma^2$, $z_dE_j$, $z_d\gamma$, $E_j\gamma$, and $E_jE_k$ for $j\neq k$), while wall terms remain additive.} For $q=7$ continuous predictors (distance, five environmental variables, and $\gamma$), the quadratic correction contributes $q(q+1)/2=28$ additional regressors. Together with the $q$ linear continuous terms already present in Eq.~\eqref{eq:pl-mw-en}, this yields $q(q+3)/2=35$ continuous regressors (excluding the intercept term $\beta_0$). Adding the two wall-count terms gives $37$ non-intercept regressors in total.

\begin{figure}[hbt!]
    \centering
    \includegraphics[width=\textwidth]{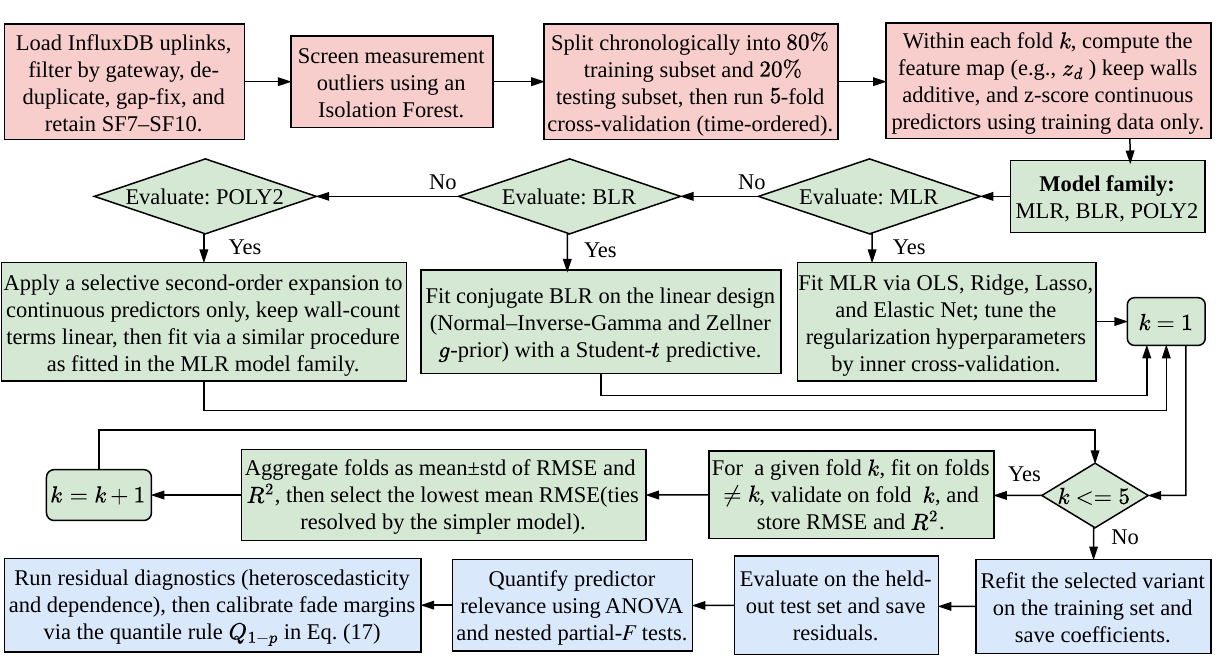}
    \caption{\CHANGES{\textbf{Model fitting pipeline.} In three phases: \textbf{(i)} database retrieval and cleaning, \textbf{(ii)} time-blocked cross-validation with within-fold preprocessing and model fitting, and \textbf{(iii)} model selection and hold-out evaluation, followed by analysis of variance (ANOVA), residual diagnostics, and fade margin calibration.}}
    \label{fig:ml_pipeline}
\end{figure}

In the data preparation pipeline shown in Fig.~\ref{fig:ml_pipeline}, raw measurements, comprising LoRaWAN-reported metadata and environmental variables, are retrieved from the InfluxDB database, cleaned upstream (including de-duplication, SF filtering, and outlier handling), and then sorted globally by timestamp. We retain SF7–SF10 to balance sensitivity and airtime \cite{grubelDenseIndoorSensor2022b}. \CHANGES{We remove measurement outliers using an Isolation Forest (contamination of $0.01$) trained on predictor-space features (excluding the response) as a data-quality screening step targeting sensor/ingestion faults. Because this procedure is unsupervised and does not use the response, it is applied prior to modeling and does not tune any model parameters.} For model fitting, we use a chronological fixed $80 \texttt{:} 20$  train/test split as deployed in \cite{gonzalez-palacioLoRaWANPathLoss2023}. \CHANGES{Within the training subset, we use $5$-fold time-ordered cross-validation using scikit-learn’s \texttt{TimeSeriesSplit} (\texttt{gap}=0). Each validation fold is a contiguous time window that follows its corresponding training window (i.e., a forward-chaining/expanding-window protocol), ensuring that no future samples appear in training for any fold. We reuse identical fold assignments across all model families, so comparisons are made on the same validation windows.} Predictors include distance $d$, frequency $f$, wall predictors $\mathbf{W}$, environmental covariates $\mathbf{E}$, and SNR $\gamma$.  We linearize distance as $10\log_{10}(d/d_0)$ and treat $20\log_{10}(f)$ as a fixed offset by subtracting it from the response during training (and adding it back for evaluation in the path loss domain).

Model comparison follows three families under the same time-ordered cross-validation protocol: \textbf{(i)} a linear MLR in Eq.~\eqref{eq:pl-mw-en} and its regularized variants fit on the linearized design (distance as $10\log_{10}(d/d_0)$ and the frequency term $20\log_{10}(f)$). Since all links are at $868\,\mathrm{MHz}$, the frequency term is a constant and is absorbed via the intercept. In multi-frequency settings, one would reintroduce the $20 \log_{10}(f)$ dependence; \textbf{(ii)} POLY2, a second-order polynomial in Eq.~\eqref{eq:pl-mw-en-quad} applied to the continuous predictors $\{z_d,E_1,\dots,E_P,\gamma\}$ while wall counts $W_k$ remain additive and linear (also fitted with the regularization specifications); \textbf{(iii)} a BLR on the linear design, reported for both a conjugate NIG prior and a Zellner $g$-prior, providing calibrated predictive uncertainty (Student-$t$ predictive) without changing the mean feature set. Out-of-sample performance is evaluated using RMSE and $R^2$ on the held-out test set. In contrast, residuals are evaluated with partial $F$ tests and \textit{Type}~II and III ANOVA, conditional on residual diagnostics (normality, independence, and variance stability).

Estimation and leakage-safe preprocessing are as follows: for penalized MLR (and for the BLR workflow, where applicable), standardization is fit within each fold using only the fold’s training indices; for POLY2, the selective quadratic expansion is applied first, followed by within-fold standardization, so that scaling reflects only the training data. \CHANGES{ Hyperparameters are selected by grid search on the same $K=5$ cross-validation folds, choosing the configuration that minimizes mean validation RMSE. We sweep Ridge $\lambda\in \mathrm{logspace}(-4,3)$ (15 points), Lasso $\lambda\in \mathrm{logspace}(-4,1)$ (15 points), and ElasticNet $\lambda\in \mathrm{logspace}(-4,1)$ (10 points) with mix $\alpha\in\{0.2,0.5,0.8\}$. For BLR, we select prior settings by the same cross-validation criterion (NIG prior scale $\Sigma_0 \in\{10^2,10^3,10^4,10^5,10^6\}$ with $a_0=b_0=10^{-2}$, and Zellner $g$-prior mode in \texttt{uip} (unit-information, $g=N$), \texttt{eb} (empirical Bayes that selects $g$ by maximizing the marginal likelihood on the training fold.) with $a_0=b_0=10^{-2}$) \cite{farawayPracticalRegressionAnova2002}.} After tuning, each selected model is refit on the full training split and evaluated once on the held-out test set. OOF residuals from the cross-validation stage are exported and used downstream for residual diagnostics and fade margin calibration (Section~\ref{subsec:fm}).

\subsection{Shadow Fading Diagnostics}
\label{subsec:shadow_method}

Our prior study on the same site reported clear deviations from Normal residuals and heavier right tails \cite{obiriStatisticalEvaluationIndoor2025a} (based on MLR). Motivated by those findings, without presupposing outcomes, we treat the residual law as an empirical object to be validated and repeat the same procedures. All steps below use the OOF residuals from the final mean specification (the second-order polynomial; Section~\ref{subsec:model_fitting}), so distributional checks and any tail modeling reflect true generalization.

We fit five families to the OOF residuals: Normal, Student’s $t$, Skew-Normal, Cauchy, and finite GMM with $K\in\{1,\ldots,5\}$ components. Parameters are estimated by maximum likelihood. For GMMs, we use the Expectation–Maximization algorithm with multiple random initializations, an identifiability constraint that orders component means for reporting, and a small variance floor to ensure numerical stability. Model choice follows a pre-specified rule to avoid hindsight bias: \textbf{(i)} compute BIC for each fitted family (and $K$ for GMMs); \textbf{(ii)} select the minimum-BIC model; \textbf{(iii)} break practical ties using the KS distance; \textbf{(iv)} when BIC or the KS distance are essentially indistinguishable, prefer the simpler family. Classical omnibus tests (D’Agostino–Pearson; Jarque–Bera) are used only as diagnostics to motivate heavier-tailed/asymmetric families when warranted, and hence they do not drive selection.

\CHANGES{For residuals $\{r_i\}_{i=1}^{N}$ and a candidate parametric density $f(r;\boldsymbol{\theta})$, we fit $\boldsymbol{\theta}$ by maximum likelihood by maximizing the log-likelihood $\mathcal{L}(\boldsymbol{\theta})$ in Eq.~\eqref{eq:log_likelihood}. We then compute AIC and BIC from the maximized log-likelihood using the standard complexity penalties ($2m$ for AIC and $m\ln(N)$ for BIC \cite{farawayPracticalRegressionAnova2002}), where $m$ is the number of estimated distribution parameters and $N$ is the sample size. Finally, we apply the KS test by comparing each fitted cumulative distribution function (CDF) with the empirical CDF of the residuals, where a smaller KS statistic indicates closer agreement over the full support.}

\begin{equation}
\mathcal{L}(\boldsymbol{\theta})=\sum_{i=1}^{N}\log\!\big(f(r_i;\boldsymbol{\theta})\big).
\label{eq:log_likelihood}
\end{equation}

Since parametric fits can mask structure when the family is misspecified, we add a model-agnostic KDE to visualize shape (core vs.\ tails) and to check for latent modality without imposing a functional form. We compute a Gaussian-kernel KDE on a fixed grid via Fast Fourier Transform (FFT) convolution (exact on the grid and computationally stable) and overlay two bandwidths: Silverman’s rule of thumb and a cross-validated log-likelihood bandwidth. We also run Hartigan’s dip test and Silverman’s critical-bandwidth test to summarize modality \cite{silvermanDensityEstimationStatistics2018}. However, the KDE is not used to produce tail quantiles; it is strictly a diagnostic to corroborate (or challenge) the parametric choice.

To check stability, mixture fits use multiple random initializations and a small variance floor, and spurious tiny-weight components are merged or discarded by the Expectation–Maximization stopping rule. Seeds are fixed for reproducibility. In the results, we report the log-likelihood, AIC, BIC, and KS test distance for each candidate, along with Q--Q plots and KDE overlays. The distribution selected by the rule above is carried forward to the tail-quantile step in Sec.~\ref{subsec:fm}, where the construction of fade margins and their uncertainty is defined.

\subsection{Fade Margin Calibration}
\label{subsec:fm}

We calibrate a fade margin that maps predictive uncertainty into a reliability buffer in the link budget of Eq.~\eqref{eq:lb_planning}. \CHANGES{For each sample $i$, let $\widehat{L}_{\ell,i}$ denote the OOF mean path loss prediction by the regressor trained without sample $i$ (equivalently, within each fold $\widehat{\mathbf{L}}=\hat{\beta}_0\mathbf{1}+\boldsymbol{\Phi}\hat{\boldsymbol{\beta}}$). The corresponding OOF error is $r_i=L_{\ell,i}-\widehat{L}_{\ell,i}$. With a fade margin $M_{\mathrm{F}}$, an outage occurs when the realized loss exceeds the predicted loss plus buffer ($L_{\ell,i}>\widehat{L}_{\ell,i}+M_{\mathrm{F}}$), i.e., $r_i>M_{\mathrm{F}}$. Thus, prescribing a target outage probability $p$ reduces to selecting $M_{\mathrm{F}}$ as an upper-tail quantile of the error distribution, consistent with classical outage-based margin setting (Q-function formulation under log-normal shadowing) \cite{goldsmithWIRELESSCOMMUNICATIONS}.} This yields the empirical fade margin estimator in Eq.~\eqref{eq:fmp}:

\CHANGES{Define the (conditional) non-exceedance probability as $\rho(M_{\mathrm{F}})\triangleq \Pr(r\le M_{\mathrm{F}})$ and the corresponding outage (exceedance) probability as $p(M_{\mathrm{F}})\triangleq \Pr(r>M_{\mathrm{F}})=1-\rho(M_{\mathrm{F}})$. In the held-out validation, we report $\hat{p}$ and its complement $\hat{\rho}=1-\hat{p}$ computed on the held-out residuals.}
{\small
\begin{equation}
\widehat{M}_{\mathrm{F},\mathrm{emp}}(p)=Q_{1-p}\!\bigl(\{r_i\}_{i=1}^{N}\bigr),
\qquad p\in(0,1)
\label{eq:fmp}
\end{equation}
}
where $Q_{1-p}(\cdot)$ is the $(1-p)$-quantile of the cross-validated residuals. When residuals are heavy-tailed or multimodal, we model the far tail with a $3$-component GMM, consistent with the residual diagnostics in Sec.~\ref{subsubsec:fit_diag_results}. \CHANGES{To guard against under-budgeting in the far tail, we define the conservative fade margin estimator $\widehat{M}_{\mathrm{F}}(p)$ as the maximum of the empirical and mixture-based tail quantiles.} For $p\le 0.02$ we also evaluate the $(1-p)$-quantile of the fitted $3$-component GMM and prescribe the conservative margin via Eq.~\eqref{eq:fm_conservative}:
{\small
\begin{equation}
\label{eq:fm_conservative}
\widehat{M}_{\mathrm{F}}(p)=\max\!\left\{\widehat{M}_{\mathrm{F},\mathrm{emp}}(p),\;
\widehat{M}_{\mathrm{F},\mathrm{GMM}}(p)\right\}.
\end{equation}
}
This conservative estimator preserves the quantile-based reliability mapping of Eq.~\eqref{eq:fmp}, while using a mixture tail model to protect the far tail \cite{papasotiriouOutdoorTHzFading2023}. Uncertainty of the empirical estimator is quantified using BCa bootstrap intervals. When short-memory dependence is detected, we use a moving-block bootstrap with the block length selected by ACF/PACF diagnostics (aggregated across devices). For mixture-based margins, we report parametric-bootstrap CIs obtained by sampling from the fitted $3$-component GMM and recomputing $Q_{1-p}$. We also summarize fold-to-fold dispersion across cross-validation splits. 

Calibration is validated on a held-out test set. \CHANGES{Let $N_{\text{test}}$ be the number of held-out test samples, and let $\mathbb{1}\{\cdot\}$ denote the indicator function (equal to $1$ if its condition holds and $0$ otherwise), with index $i\in\{1,\ldots,N_{\text{test}}\}$.} For a prescribed $\widehat{M}_{\mathrm{F}}(p)$, we compute the achieved outage using Eq.~\eqref{eq:achieved_outage}:
{\small
\begin{equation}
\label{eq:achieved_outage}
\hat{p}=\frac{1}{N_{\text{test}}}\sum_{i=1}^{N_{\text{test}}} \mathbb{1}\{L_{\ell,i} > \widehat{L}_{\ell,i} + \widehat{M}_{\mathrm{F}}(p)\}.
\end{equation}
}
\CHANGES{Since $L_{\ell,i}$ is observed only for received packets (Eq.~\eqref{eq:pl_from_rssi}), the reported achieved reliability is $\hat{\rho}=1-\hat{p}=\Pr(r\le \widehat M_{\mathrm F}(p)\mid \text{received})$, i.e., a conditional non-exceedance on received packets.}  Sweeping $p\in\{0.05,0.02,0.01\}$, we plot the achieved reliability $\hat{\rho}$ against the prescribed margin and overlay target iso-lines, where alignment indicates reliability-correct calibration on unseen data. Finally, we report $M_{\mathrm{F},99} \equiv M_{\mathrm{F}}(0.01)$ with $95\%$ CIs, compare against a $10\,\mathrm{dB}$ fixed heuristic margin, \CHANGES{adopted from \cite{barriquelloFundamentalsWirelessCommunication2017},} and include achieved conditional reliability on received packets $\hat{\rho}$ versus the fade margin calibration visualization.

\section{Results and Analysis}
\label{sec:results}

\CHANGES{This section follows the pipeline in Fig.~\ref{fig:ml_pipeline} to: \textbf{(i)} report fitting predictive performance, \textbf{(ii)} interpret the fitted mean, \textbf{(iii)} summarize residual diagnostics, and \textbf{(iv)} use them to support fade margin calibration with held-out validation.}

\subsection{Regression}\label{subsec:regression}

\begin{table}[hbt!]
\centering

\caption{\CHANGES{\textbf{Comparative performance across the model families (RMSE and coefficient of determination ($\bm{R^2}$)).} We report the best configuration for each model family (OLS/Ridge/Lasso/ElasticNet), selected by the OOF mean RMSE, with $\pm$ denoting standard deviation across folds. Within-family OOF RMSE ranges are as follows: \textbf{(i)} COST 231 MWM $\le 0.01\,\mathrm{dB}$, \textbf{(ii)} MLR $\le 0.01\,\mathrm{dB}$, \textbf{(iii)} BLR $\approx 0\,\mathrm{dB}$, and \textbf{(iv)} POLY2 $\le 0.09\,\mathrm{dB}$.}}

\label{tab:perf_comparison}
\setlength{\tabcolsep}{3pt}
\begin{tabularx}{\columnwidth}{@{}>{\centering\arraybackslash}p{2.0cm} *{2}{Y} *{2}{Y}@{}}
\toprule
\multirow{2}{*}{\textbf{Model Family}} &
\multicolumn{2}{c}{\textbf{RMSE} ($\bm{\mathrm{dB}}$)} &
\multicolumn{2}{c}{\boldmath$\bm{R^2}$} \\
\cmidrule(lr){2-3}\cmidrule(lr){4-5}
& \textbf{Cross-validated (OOF)} & \textbf{Test (held-out)}
& \textbf{Cross-validated (OOF)} & \textbf{Test (held-out)} \\
\midrule
COST 231 \cite{europeancommissionCOSTAction2311999}   & $10.9739 \pm 1.0518$ & $12.0704$ & $0.6533$ & $0.5890$ \\
MLR   & $8.2333 \pm 0.5577$ & $8.4845$ & $0.8061$ & $0.7969$ \\
BLR   & $8.2416\pm 0.5827$ & $8.4534$ & $0.8055$ & $0.7984$ \\
\textbf{POLY2} & $\bm{7.3761 \pm 0.6048}$ & $\bm{7.7708}$ & $\bm{0.8439}$ & $\bm{0.8296}$ \\
\bottomrule
\end{tabularx}
\end{table}

\CHANGES{From Table~\ref{tab:perf_comparison}, the empirical COST 231 MWM, included as a structural-only reference (distance and wall obstructions), exhibits substantially weaker generalization ($\mathrm{RMSE}\approx 12.07\,\mathrm{dB}$, $R^2\approx 0.59$), indicating that structural descriptors alone underfit the variability present in the measurements.} The extended linear baselines (MLR and BLR) are essentially indistinguishable under the time-blocked protocol, achieving a cross-validated RMSE of $\approx 8.24\,\mathrm{dB}$ with a test RMSE of $\approx 8.47\,\mathrm{dB}$ and $R^2\approx 0.80$ for both cases. This stability suggests the linear regime is well-conditioned for the available predictors, and that any multicollinearity among environmental covariates does not materially affect predictive accuracy \cite{geronHandsonMachineLearning}. While BLR does not improve point-error metrics relative to MLR, it retains practical value by providing calibrated posterior uncertainty (e.g., coefficient credible intervals), which is useful when link variability is modulated by occupancy and ventilation dynamics. In contrast, the second-order polynomial substantially improves generalization, reducing the cross-validated RMSE to $7.3761\,\mathrm{dB}$ ($\approx 10.4\%$) and, respectively, improving $R^2$ to $0.8439$, with consistent gains on the held-out test. This indicates that second-order interaction terms capture nonlinear dependencies not captured by purely linear models, thereby improving on the MLR baseline.

\begin{table}[hbt!]
\centering
\caption{\CHANGES{\textbf{Predictor coefficients for the linear model (OLS) with HC3 $\bm{95\%}$ confidence intervals (CIs).} HC3 refers to the MacKinnon--White heteroskedasticity-consistent covariance estimator.}}
\label{tab:modelcoeffs}
\setlength{\tabcolsep}{6pt}
\begin{tabular*}{\linewidth}{@{\extracolsep{\fill}}
  >{\centering\arraybackslash}p{3.5cm}
  >{\centering\arraybackslash}p{2.0cm}
  >{\centering\arraybackslash}p{2.0cm}
  Z{+1.4}
  Z{+1.4}
  Z{+1.4}
  @{}}
\toprule
\multicolumn{1}{c}{\multirow{2}{*}{\textbf{Variable}}} &
\multicolumn{1}{c}{\multirow{2}{*}{\textbf{Unit}}} &
\multicolumn{1}{c}{\multirow{2}{*}{\textbf{Coefficient}}} &
\multicolumn{1}{c}{\multirow{2}{*}{\textbf{Estimate}}} &
\multicolumn{2}{c}{\textbf{95\% CIs (HC3) ($\bm{\mathrm{dB}}$)}} \\
\cmidrule(lr){5-6}
& & & & \multicolumn{1}{c}{\textbf{lower}} & \multicolumn{1}{c}{\textbf{upper}} \\
\midrule
Intercept             & $\mathrm{dB}$                         & $\beta_0$                   & 2.31    & 1.80    & 2.81 \\
Path loss exponent    & --                                    & $n$                         & 3.87    & 3.86    & 3.88 \\
Brick wall loss       & $\mathrm{dB}$                         & $\omega_{\mathrm{brick}}$   & 6.83    & 6.81    & 6.85 \\
Wood partition loss   & $\mathrm{dB}$                         & $\omega_{\mathrm{wood}}$    & 1.98    & 1.96    & 1.99 \\
Carbon dioxide        & $\mathrm{dB/ppm}$                     & $\varepsilon_{\mathrm{C}}$  & -0.0024 & -0.0025 & -0.0022 \\
Relative humidity     & $\mathrm{dB/\%}$                      & $\varepsilon_{\mathrm{RH}}$ & -0.0917 & -0.0938 & -0.0897 \\
Particulate matter    & $\mathrm{dB}/(\mu\mathrm{g/m^3})$     & $\varepsilon_{\mathrm{PM}}$ & -0.0953 & -0.1011 & -0.0895 \\
Barometric pressure   & $\mathrm{dB/hPa}$                     & $\varepsilon_{\mathrm{BP}}$ & -0.0080 & -0.0094 & -0.0066 \\
Temperature           & $\mathrm{dB}/\si{\degreeCelsius}$     & $\varepsilon_{\mathrm{T}}$  & -0.1410 & -0.1443 & -0.1378 \\
SNR scaling           & --                                    & $k_{\gamma}$                & -2.0344 & -2.0426 & -2.0262 \\
\bottomrule
\end{tabular*}
\end{table}

\CHANGES{For Table~\ref{tab:modelcoeffs}, we report the OLS coefficients as a representative linear fit together with $95\%$ confidence intervals computed using the MacKinnon--White heteroskedasticity-consistent covariance estimator (HC3; leverage-adjusted) to quantify coefficient uncertainty under potential heteroskedasticity. Across the frequentist linear variants, Ridge and Lasso yield near-identical estimates to OLS, with differences at most $0.01$ for the dominant geometric terms (path loss exponent and wall losses) and around $10^{-3}$ or smaller for the environmental slopes. Likewise, the two BLR specifications yield the same point estimates to numerical precision, consistent with a data-dominant likelihood. Their practical advantage is calibrated posterior uncertainty (e.g., coefficient credible intervals) rather than lower point-error. ElasticNet can induce stronger shrinkage on the correlated environmental block, effectively redistributing weights among covariates without materially changing predictive performance (Table~\ref{tab:perf_comparison}). For interpretability, we therefore summarize the linear regime using the OLS coefficients.}

The fitted intercept reflects fixed offsets (including free space path loss $(f,1\,\mathrm{m})$ at $868\,\mathrm{MHz}$ and centering), so we do not interpret $\beta_{0}$ physically. The path loss exponent ($n \approx 3.87$) is consistent with established indoor LoRaWAN ranges and reflects rapid decay in a multi-obstruction office environment \cite{goldsmithWIRELESSCOMMUNICATIONS, azevedoCriticalReviewPropagation2024}. The estimated obstruction losses show the expected material contrast, with brick inducing about $6.83\,\mathrm{dB}$ per wall compared with about $1.98\,\mathrm{dB}$ for wooden partitions, in line with reported attenuation differences for building materials around $868\,\mathrm{MHz}$ \cite{ruddBuildingMaterialsPropagation2014}. Moreover, the environmental terms retain the same sign pattern, whereas the SNR scaling was strongly negative ($k_{\gamma} \approx -2.03$). In contrast to typical outdoor regression settings, where meteorological variables are often treated as direct propagation modifiers \cite{gonzalez-palacioLoRaWANPathLoss2023}, indoors, these quantities should be interpreted primarily as conditional state proxies rather than causal attenuation terms. \CHANGES{Elevated CO\textsubscript{2} and humidity co-vary with occupancy and HVAC scheduling, i.e., regimes in which the indoor channel is time-varying due to human motion, door usage, and airflow control. Consequently, their fitted signs reflect a correlation with building-operation states that shift the conditional link budget after controlling for distance and wall counts, rather than radiative absorption in the strict physical sense \cite{grubelDenseIndoorSensor2022b}.} The SNR term is expected to be strongly informative because it acts as an instantaneous demodulation-margin indicator tied to received power and noise; including it captures fast channel and interference conditions and reduces residual variance, but it should not be interpreted as an independent physical loss mechanism \cite{rappaportWirelessCommunicationsPrinciples2002}. Collectively, the environmental block functions as an implicit low-dimensional channel-state indicator, supporting environment-aware reliability modeling for indoor IoT operation at $868\,\mathrm{MHz}$.

\CHANGES{For the best-performing nonlinear model (POLY2, Lasso-selected), the intercept absorbs fixed offsets as well as centering and standardization effects induced by the polynomial expansion; accordingly, we do not attach direct physical meaning to it. The distance dependence is captured primarily by curvature in the distance term, while the linear distance contribution is strongly regularized.}
We verified numerically that the net gradient $\partial \widehat{L}_{\ell,\mathrm{poly}}/\partial z_d$ remains positive across the evaluated distance range when conditioned on the observed covariates. Wall-loss effects remain physically consistent with material-dependent attenuation, in line with sub-GHz measurement evidence \cite{sebastianDielectricMaterialsWireless2010}. Most first-order environmental slopes (e.g., $\varepsilon_{\mathrm{C}},\,\varepsilon_{\mathrm{RH}}$) are shrunk toward zero under regularization, whereas a small set of interaction terms involving distance, humidity, and link-state indicators remains nonzero. This pattern supports a conditional channel-state interpretation that the retained environment--distance and environment--SNR couplings likely act as proxies for occupancy/HVAC-regime changes and other latent dynamics rather than direct atmospheric absorption at $868\,\mathrm{MHz}$. A first-order term with mild curvature dominates link-state dependence. Finally, removing the SNR predictor increases RMSE while preserving the qualitative role of the retained environmental interactions.

\begin{table}[hbt!]
\centering
\caption{\CHANGES{\textbf{Additive MLR significance tests.} Residual degrees of freedom (df) are $\n{1663617}$. Partial $\eta^2$ is computed as $F\,\mathrm{df}_1/(F\,\mathrm{df}_1+\mathrm{df}_2)$, where $\mathrm{df}_1$ and $\mathrm{df}_2$ denote the numerator and denominator degrees of freedom, respectively; for Panel~A, $\mathrm{df}_1=1$. HC3 refers to the MacKinnon--White heteroskedasticity-consistent covariance estimator.}}
\label{tab:mlr_significance}
\setlength{\tabcolsep}{6.5pt}
\renewcommand{\arraystretch}{1.05}

\begin{tabular*}{\columnwidth}{@{\extracolsep{\fill}}@{}
  >{\centering\arraybackslash}p{2.8cm}
  >{\centering\arraybackslash}p{1.5cm}
  >{\centering\arraybackslash}p{2.4cm}
  Z{+7.3}
  c
  Z{1.4}
  c
  @{}}
\toprule
\multicolumn{7}{c}{\textbf{Panel A: HC3-robust Type II ANOVA-style per-predictor tests}}\\
\midrule
\textbf{Variable} & \textbf{Unit} & \textbf{Coefficient} &
\multicolumn{1}{c}{\textbf{$\bm{F}$}} &
\textbf{$\bm{p}$} &
\multicolumn{1}{c}{\textbf{partial $\bm{\eta^2}$}} &
\textbf{Sign} \\
\midrule
Path loss exponent       & --                                & $n$                         & 439732.636   & \pmin & 0.2091 & $+$ \\
Brick wall loss          & $\mathrm{dB}$                     & $\omega_{\mathrm{brick}}$   & 318053.020   & \pmin & 0.1605 & $+$ \\
Wood partition loss      & $\mathrm{dB}$                     & $\omega_{\mathrm{wood}}$    &  55431.275   & \pmin & 0.0322 & $+$ \\
Carbon dioxide           & $\mathrm{dB/ppm}$                 & $\varepsilon_{\mathrm{C}}$  &   1841.148   & \pmin & 0.0011 & $-$ \\
Relative humidity        & $\mathrm{dB/\%}$                  & $\varepsilon_{\mathrm{RH}}$ &   7743.697   & \pmin & 0.0046 & $-$ \\
Particulate matter       & $\mathrm{dB}/(\mu\mathrm{g/m^3})$ & $\varepsilon_{\mathrm{PM}}$ &   1025.772   & \pmin & 0.0006 & $-$ \\
Barometric pressure      & $\mathrm{dB/hPa}$                 & $\varepsilon_{\mathrm{BP}}$ &    126.342   & \pmin & 0.0001 & $-$ \\
Temperature              & $\mathrm{dB}/\si{\degreeCelsius}$ & $\varepsilon_{\mathrm{T}}$  &   7228.963   & \pmin & 0.0043 & $-$ \\
SNR                      & --                                & $k_{\gamma}$                & 236234.977   & \pmin & 0.1243 & $-$ \\
\midrule

\addlinespace[0.8em]

\multicolumn{7}{c}{\textbf{Panel B: Nested block comparisons (OLS residual sum of squares partial-$F$)}}\\
\midrule
\multicolumn{2}{c}{\textbf{Comparison}}
& \textbf{df$_1$, df$_2$}
& \multicolumn{1}{c}{\textbf{$\bm{F}$}}
& \textbf{$\bm{p}$}
& \multicolumn{1}{c}{\textbf{partial $\bm{\eta^2}$}} & \\
\multicolumn{2}{l}{Structure $\rightarrow$ +Environmentals} & (\n{5},\,\n{1663618}) & 3298.886    & \pmin & 0.0098 & \\
\multicolumn{2}{l}{+Environmentals $\rightarrow$ +SNR}      & (\n{1},\,\n{1663617}) & 1258588.250 & \pmin & 0.4307 & \\
\bottomrule
\end{tabular*}
\end{table}

\subsection{Analysis of Variance (ANOVA)}
\label{sec:anova}

For consistency across models, we report \textit{Type}~II and \textit{Type}~III ANOVA-style $F$ tests for the additive MLR baseline using the same HC3-robust covariance estimator as in Table~\ref{tab:modelcoeffs}, and compare nested predictor blocks via classical OLS residual sum of squares partial-$F$ tests (structure, then +environmental variables, then +SNR). \CHANGES{Since the sample size is very large ($N\approx 1.66\times10^6$), the associated $p$-values become numerically saturated (often effectively $0$). Thus, statistical significance here should be read as detectability rather than practical magnitude \cite{langsrudANOVAUnbalancedData2003}. We therefore interpret effects primarily via partial $\eta^2$ (an $F$-derived partial effect-size index that coincides with the usual ANOVA partial-variance share under homoskedastic OLS), which is bounded and comparable across predictors.}

The resulting relative ordering is summarized in Table~\ref{tab:mlr_significance}, from which we conclude that: \textbf{(i)} Per-predictor (Panel A): Structural terms dominate, with the path loss exponent contributing most (partial $\eta^2 \approx 0.21$), followed by brick-wall loss. The SNR term is of comparable magnitude(partial $\eta^2 \approx 0.12$) and exceeds the wood-partition term, while the environmental covariates are individually modest but directionally consistent once structure is controlled. The ranking and signs are stable under \textit{Type}~III (discussed next). \textbf{(ii)} Block gains (Panel B): In Panel B, adding the five environmental covariates after structure yields a statistically detectable but modest reduction in residual sum of squares ((partial $\eta^2 \approx 0.0098$), whereas adding SNR afterward delivers the dominant improvement (partial $\eta^2 \approx 0.4307$), indicating that SNR carries substantial non-redundant link-state information beyond structure and microclimate. \CHANGES{This is consistent with the gap between COST 231 MWM and MLR results in Table~\ref{tab:perf_comparison}.  Since the MLR specification includes only main effects (no interactions), \textit{Type}~III ANOVA yields the same tests as \textit{Type}~II for all predictors, and hence the effect-size ordering is insensitive to ANOVA type. \textit{Type}~III ANOVA additionally reports an intercept test, which we do not interpret physically because it absorbs fixed offsets (e.g., the constant frequency term).}

\begin{table}[hbt!]
\centering
\caption{\CHANGES{\textbf{Partial-$\bm{F}$ tests.} Left: nested block additions, and right: drop-one (second-order polynomial; train set).}}
\setlength{\tabcolsep}{7.5pt}
\label{tab:partial_f}
\begin{tabular}{
  l
  Z{2.0}
  Z{+6.0}
  c
  Z{1.3}
  l
  Z{2.0}
  Z{+6.0}
  Z{1.3}
}
\toprule
\multicolumn{5}{c}{\textbf{Nested additions}} & \multicolumn{4}{c}{\textbf{Drop-one versus full}} \\
\cmidrule(lr){1-5}\cmidrule(lr){6-9}
\textbf{Block}
  & \multicolumn{1}{c}{\textbf{$\bm{\Delta df}$}}
  & \multicolumn{1}{c}{\textbf{$\bm{F}$}}
  & \textbf{$\bm{p}$}
  & \multicolumn{1}{c}{\textbf{$\bm{\eta^2}$}}
  & \textbf{Dropped}
  & \multicolumn{1}{c}{\textbf{$\bm{\Delta df}$}}
  & \multicolumn{1}{c}{\textbf{$\bm{F}$}}
  & \multicolumn{1}{c}{\textbf{$\bm{\eta^2}$}} \\
\midrule
$D$ ($z_d, z_d^2$)  &  2  & 184141 & $<10^{-16}$ & 0.181 & $W$ ($\mathbf{w}$)         &  2 & 198617 & 0.193 \\
$E$ ($\mathbf{e}$)     & 10  &   1346 & $<10^{-16}$ & 0.008 & $D$ ($z_d, z_d^2$)  &  2 &  21039 & 0.025 \\
$S$ ($\gamma$, $\gamma^2$)  &  2  & 878961 & $<10^{-16}$ & 0.514 & $E$ ($\mathbf{e}$)     & 10 &    307 & 0.002 \\
$X$ (cross-terms in $\mathbf{u}$)  & 21  &   8959 & $<10^{-16}$ & 0.102 & $S$ ($\gamma$, $\gamma^2$)  &  2 & 143520 & 0.147 \\
                    &     &        &             &       & $X$ (cross-terms in $\mathbf{u}$)  & 21 &   8959 & 0.102 \\
\bottomrule
\end{tabular}
\end{table}

\CHANGES{To dissect predictor contributions in the polynomial model, we conducted partial-$F$ tests on nested blocks denoted as: $W$ (walls), $D$ (distance), $E$ (environment-only), $S$ (SNR), and $X$ (all interactions of $z_d$, $\gamma$ and $\mathbf{e}$, excluding squares) (Table~\ref{tab:partial_f}).} In nested additions, distance provides a strong initial gain (partial $\eta^2 = 0.181$), SNR contributes the largest incremental improvement (partial $\eta^2 = 0.514$), and interactions add a further non-trivial component (partial $\eta^2 = 0.102$). In contrast, the environment-only block is modest in isolation (partial $\eta^2 = 0.008$). In the drop-one tests, we remove the indicated term block and compare the reduced model to the full model. Removing walls or SNR incurs the steepest penalties (partial $\eta^2 = 0.193$ and $0.147$, respectively), confirming the primacy of structural attenuation and instantaneous link state. Omitting interactions also causes a measurable loss (partial $\eta^2 = 0.102$), indicating that nonlinear couplings refine fidelity beyond additive effects. Therefore, the environment block contributes most meaningfully when expressed through interactions rather than as purely additive slopes, supporting environment-aware reliability tuning without sacrificing coverage.

\subsection{Shadow Fading Analysis}
\label{subsec:residdist}

\subsubsection{Model Fit Diagnostics}
\label{subsubsec:fit_diag_results}

\begin{table}[hbt!]
\centering
\caption{\CHANGES{\textbf{Distributional fit diagnostics on the second-order polynomial regression residuals ($\bm{N=1663627}$).}}}
\label{tab:distfits}
\setlength{\tabcolsep}{3pt}
\begin{tabularx}{\columnwidth}{@{}>{\centering\arraybackslash}p{2.8cm} *{4}{Y}@{}}
\toprule
\textbf{Distribution} & \multicolumn{1}{c}{\textbf{Log-Likelihood}} & \multicolumn{1}{c}{\textbf{AIC}} &
\multicolumn{1}{c}{\textbf{BIC}} & \multicolumn{1}{c}{\textbf{KS test}} \\
\midrule
Normal & -5669116.41 & 11338236.83 & 11338261.48 & 0.0603 \\
Skew--Normal & -5627994.19 & 11255994.38 & 11256031.35 & 0.0483 \\
GMM (K=3) & -5513071.13 & 11026158.25 & 11026256.85 & 0.0108 \\
Cauchy & -5729734.65 & 11459473.30 & 11459497.95 & 0.0816 \\
$t$-Distribution & -5537405.70 & 11074817.41 & 11074854.38 & 0.0279 \\
\bottomrule
\end{tabularx}
\end{table}

\CHANGES{Residual diagnostics for the second-order polynomial model reject Normality decisively (Omnibus/D’Agostino statistic $\approx 3.49\times 10^{5}$ and Jarque--Bera statistic $\approx 3.08\times 10^{6}$, both $p<10^{-16}$), with a right-skewed and heavy-tailed shape (skewness $\approx 0.765$, excess kurtosis $\approx 6.488$). Autocorrelation is not a dominant concern at this resolution (Durbin--Watson $\approx 2.022$). Even with a strong predictive fit (OOF $R^2 \approx 0.844$), a single Gaussian law is inadequate, consistent with our earlier baseline MLR analysis \cite{obiriStatisticalEvaluationIndoor2025a}.} As summarized in Table~\ref{tab:distfits}, a three-component Gaussian mixture provides the best residual description, achieving the lowest BIC and the smallest KS statistic. Skew-Normal improves over Normal but still under-represents the right tail, while the $t$ distribution offers an intermediate fit. This residual-tail characterization forms the basis of the fade margin calibration in Sec.~\ref {subsec:fm}.

\subsubsection{Parametric Residual Distributions}
\label{subsubsec:parametric_results}

\begin{figure}[hbt!]
    \centering
    \includegraphics[width=\textwidth]{qq_residual_plots.png}
    
    \caption{\label{fig:qqplots} \CHANGES{\textbf{Parametric residual distribution diagnostics.} Q--Q plots of the second-order polynomial OOF residuals against candidate parametric families are shown in \textbf{(a)}--\textbf{(e)}, and \textbf{(f)} shows the residual histogram with fitted densities. The selected three-component GMM best captures the sharp central mass and heavy tails, comprising two dominant narrow components and a lower-weight broad component.}}
    
\end{figure}

Table~\ref{tab:distfits} reports the KS statistics and information criteria for the candidate families. Due to the large value of $N$, KS $p$-values saturate, so we compare KS statistics and BIC. Table~\ref{tab:distfits} and Fig.~\ref{fig:qqplots} support the following observations: \textbf{(i)} Normal underestimates tail mass, with KS $\approx 0.0603$ (Fig.~\ref{fig:qqplots}(a)). \textbf{(ii)} Skew--Normal captures asymmetry better (shape $\approx 1.60$) but still deviates in the extreme quantiles (KS $\approx 0.0483$; Fig.~\ref{fig:qqplots}(b)). \textbf{(iii)} A Gaussian mixture with $K{=}3$ achieves the best parsimony--fit trade-off (lowest BIC $\approx 1.1026\times 10^{7}$; KS $\approx 0.0108$) and closely tracks both the core and the tails (Fig.~\ref{fig:qqplots}(c)). \CHANGES{The fitted mixture comprises two dominant narrow components and a lower-weight broad component, consistent with heterogeneous indoor shadowing states \cite{yaoHierarchicalPositioningModel2025}.} \textbf{(iv)} Cauchy over-weights extremes and misrepresents the bulk (KS $\approx 0.0816$; Fig.~\ref{fig:qqplots}(d)). \textbf{(v)} Student’s $t$ thickens tails (degrees of freedom $\nu \approx 4.29$) and improves fit relative to Normal, but remains intermediate overall (KS $\approx 0.0279$; Fig.~\ref{fig:qqplots}(e)). Compared with MLR-based site analysis in \cite{obiriStatisticalEvaluationIndoor2025a} that favored a higher mixture order, the polynomial mean reduces residual heterogeneity to $K{=}3$, indicating that mild nonlinear couplings are absorbed. At the same time, a compact multimodal shadow-fading structure remains. This selected residual law is carried forward to the fade margin calibration in Sec.~\ref{subsec:fm}.

\subsubsection{Nonparametric Density and Modality}
\label{subsubsec:kde_results}

\begin{figure}[hbt!]
  \centering
  \includegraphics[width=\linewidth]{residual_kde__and__modes_vs_bw__side_by_side.png}

  \caption{\CHANGES{\textbf{Nonparametric residual density and modality diagnostics.} \textbf{(a)} Residual density with kernel density estimation (KDE) overlays using Silverman’s rule ($h\approx 0.307\,\mathrm{dB}$) and a cross-validated log-likelihood (CV-LL) bandwidth ($h\approx 0.975\,\mathrm{dB}$). \textbf{(b)} Estimated mode count versus bandwidth on a logarithmic sweep using prominence-based peak detection.}}
  
  \label{fig:kde_modes}
\end{figure}

As shown in Fig.~\ref{fig:kde_modes}(a), both KDE views indicate a sharp central mass with a mild shoulder and a slightly heavier right tail. The narrower bandwidth ($h\approx 0.307\,\mathrm{dB}$) reveals local structure, whereas the broader bandwidth ($h\approx 0.975\,\mathrm{dB}$) merges minor bumps into an effectively unimodal core \cite{silvermanDensityEstimationStatistics2018}. Formal modality diagnostics are consistent with this picture. Hartigan’s dip test rejects strict unimodality at the native scale ($p<10^{-16}$), but the bandwidth required to enforce unimodality is modest since Silverman’s critical bandwidth is $h^\ast\approx 0.466\,\mathrm{dB}$ (smoothed-bootstrap $p\approx 1.00$ for $H_0\!:\leq 1$ mode). The mode-count curve in Fig.~\ref{fig:kde_modes}(b) mirrors this behavior, with many spurious peaks at very small $h$ that rapidly collapse toward a single mode as smoothing increases, reaching unimodality around $h\sim 0.5\,\mathrm{dB}$. Together with the BIC-selected $K{=}3$ GMM (Table~\ref{tab:distfits}), the evidence supports a near-unimodal core with a low-probability heavy-tail regime, motivating mixture-based tail calibration for fade margins while using KDE strictly as a diagnostic view of residual shape.

\begin{table}[hbt!]
\centering

\caption{\CHANGES{\textbf{Group-wise OOF residual location and dispersion diagnostics.} MAD denotes the median absolute deviation, where $\sigma \approx 1.4826\,\mathrm{MAD}$ under Gaussian residuals \cite{rousseeuwAlternativesMedianAbsolute1993}; $\epsilon^2_{\mathrm{KW}}$ is the Kruskal--Wallis effect size for location/rank differences, and $\eta^2_{\mathrm{BF}}$ is the Brown--Forsythe effect size for dispersion based on the median-centered Levene test.}}

\label{tab:resid_group_tests}

\setlength{\tabcolsep}{6pt}
\renewcommand{\arraystretch}{1.05}

\begin{tabular*}{\linewidth}{@{\extracolsep{\fill}}@{} l l Z{7.0} Z{+1.2} Z{1.2} Z{1.4} Z{1.4} @{}}

\toprule
\textbf{Partition} &
\textbf{Group} &
\multicolumn{1}{c}{\textbf{$\bm{N_{\mathrm{grp}}}$}} &
\multicolumn{1}{c}{\textbf{Median ($\bm{\mathrm{dB}}$)}} &
\multicolumn{1}{c}{\textbf{MAD ($\bm{\mathrm{dB}}$)}} &
\multicolumn{1}{c}{\textbf{$\bm{\epsilon^2_{\mathrm{KW}}}$}} &
\multicolumn{1}{c}{\textbf{$\bm{\eta^2_{\mathrm{BF}}}$}} \\
\midrule

\multirow{2}{*}{LoS vs NLoS}
  & LoS  & 277295  & -1.98 & 4.14
  & \multicolumn{1}{c}{\multirow{2}{*}{\num{0.0111}}}
  & \multicolumn{1}{c}{\multirow{2}{*}{\num{0.0007}}} \\
  & NLoS & 1386332 & -0.27 & 3.97 & & \\

\midrule

\multirow{3}{*}{CO\textsubscript{2} terciles}
  & Low  & 557015 & -0.82 & 3.95
  & \multicolumn{1}{c}{\multirow{3}{*}{\num{0.0008}}}
  & \multicolumn{1}{c}{\multirow{3}{*}{\num{0.0001}}} \\
  & Mid  & 543064 & -0.30 & 4.06 & & \\
  & High & 563548 & -0.57 & 4.06 & & \\

\bottomrule
\end{tabular*}
\end{table}

\CHANGES{We also tested whether coarse context partitions explain the residual heterogeneity using full-sample Kruskal--Wallis (location) and Brown--Forsythe (scale) tests (Table~\ref{tab:resid_group_tests}). Although differences are statistically detectable at this sample size, the corresponding effect sizes are small: LoS/NLoS accounts for only about $1\%$ of rank-based location variability and well below $0.1\%$ of dispersion, while CO\textsubscript{2} terciles are smaller still. These partitions, therefore, explain little of the residual geometry, so the mixture-protected tail treatment remains justified for far-tail fade margin calibration rather than being an artifact of stratifying by LoS/NLoS or CO\textsubscript{2} alone.}

\newcommand{\Shead}[1]{\multicolumn{1}{c}{\textbf{#1}}}
\begin{table}[hbt!]
\centering
\setlength{\tabcolsep}{8pt}
\renewcommand{\arraystretch}{1.05}
\caption{\CHANGES{\textbf{Fade margin calibration from out-of-fold (OOF) residuals and held-out validation.} The table summarizes the calibrated fade margin $M_{\mathrm{F}}$ for target outage levels $100p(\%)$, based on either empirical quantiles or GMM-tail estimation, alongside $95\%$ confidence intervals and the achieved held-out reliability $\hat{\rho}$.}}
\label{tab:fm_calibration}
\begin{tabular*}{\linewidth}{@{\extracolsep{\fill}}@{}
  c
  S[table-format=1.0]
  c
  S[table-format=2.2]
  S[table-format=2.2]
  S[table-format=2.2]
  S[table-format=1.4]
  @{}}
  
\toprule
\multicolumn{1}{c}{\multirow{2}{*}{\textbf{Model}}} &
\multicolumn{1}{c}{\multirow{2}{*}{\textbf{$\bm{100p(\%)}$}}} &
\multicolumn{1}{c}{\multirow{2}{*}{\textbf{Estimator}}} &
\multicolumn{1}{c}{\multirow{2}{*}{\textbf{$\bm{M_{\mathrm{F}}}$ ($\bm{\mathrm{dB}}$)}}} &
\multicolumn{2}{c}{\textbf{$\bm{95\%}$ CIs ($\bm{\mathrm{dB}}$)}} &
\multicolumn{1}{c}{\multirow{2}{*}{\textbf{Achieved reliability $\bm{\hat{\rho}}$}}}\\
\cmidrule(lr){5-6}
& & & & \Shead{lower} & \Shead{upper} & \\
\midrule

COST 231 MWM & 5 & Empirical & 12.09 & 12.08 & 12.15 & 0.9544 \\
COST 231 MWM & 2 & Empirical & 44.61 & 44.40 & 44.61 & 0.9705 \\
COST 231 MWM & 1 & Empirical & 60.76 & 60.75 & 61.07 & 0.9898 \\
\midrule
BLR   & 5 & Empirical & 13.18 & 13.12 & 13.24 & 0.9565 \\
BLR   & 2 & GMM tail  & 21.86 & 21.76 & 21.96 & 0.9810 \\
BLR   & 1 & Empirical & 27.79 & 27.40 & 28.13 & 0.9866 \\
\midrule
MLR   & 5 & Empirical & 12.99 & 12.93 & 13.05 & 0.9559 \\
MLR   & 2 & GMM tail  & 21.59 & 21.48 & 21.70 & 0.9801 \\
MLR   & 1 & Empirical & 28.05 & 27.65 & 28.44 & 0.9865 \\
\midrule
POLY2 & 5 & Empirical & 11.96 & 11.89 & 12.01 & 0.9608 \\
POLY2 & 2 & GMM tail  & 19.65 & 19.55 & 19.76 & 0.9793 \\
POLY2 & 1 & GMM tail  & 25.73 & 25.61 & 25.85 & 0.9856 \\
\bottomrule
\end{tabular*}
\end{table}

\begin{figure}[hbt!]
    \centering
    \includegraphics[width=\linewidth]{rho_fm_with_delta_savings.png}
    \caption{\CHANGES{\textbf{Fade margin calibration on held-out test set.} (a) Held-out achieved reliability $\hat{\rho}$ versus fade margin with calibrated operating points $\widehat{M}_{\mathrm{F}}(p)$ for $p\in\{5,2,1\}\%$ and $95\%$ confidence intervals. The COST 231 MWM is shown for context, but its calibrated margins for $p\le 2\%$ exceed the plotted range (Table~\ref{tab:fm_calibration}). (b) Fade margin differences $\Delta M_{\mathrm{F}}(p)$ ($M_{\mathrm{F},A}(p)-M_{\mathrm{F},B}(p)$) among BLR/MLR/POLY2 models at matched outage targets (COST 231 MWM omitted for scale)}}
    \label{fig:rho_vs_fm}
\end{figure}

\subsection{Fade margin calibration}
\label{subsec:fm_results}

Following the procedure in Sec.~\ref{subsec:fm}, we estimate $M_{\mathrm{F}}(p)$ from the residuals and validate calibration on held-out data via the achieved reliability $\hat{\rho}$ (computed on received packets only as a conditional non-exceedance) at the prescribed $\widehat{M}_{\mathrm{F}}(p)$. Table~\ref{tab:fm_calibration} reports the calibrated margins, bootstrap $95\%$ confidence intervals, and held-out achieved reliability $\hat{\rho}$. As shown in Fig.~\ref{fig:rho_vs_fm}\,(a), the calibrated operating points lie close to the target iso-lines, and the held-out achieved reliability is near the nominal levels \CHANGES{($\leq \sim 1.1\%$ absolute across all targets including $95\%$, and $\leq \sim 0.5\%$ for the stricter outage targets ($p \leq 2\%$))}, which is a considerably reliable calibration under our proposed protocol. Fig.~\ref{fig:rho_vs_fm}\,(b) summarizes the comparison as fade margin differences $\Delta M_{\mathrm{F}}(p)$, where positive values mean a model requires more margin. The linear baselines (BLR/MLR) are essentially tied (within $\approx 0.3\,\mathrm{dB}$), while POLY2 steadily lowers the required margin, with the benefit growing as $p$ tightens. \CHANGES{At $p=1\%$, the reduction is $\approx 2.32\,\mathrm{dB}$, roughly one SF step ($\approx 3\,\mathrm{dB}$ transmit power). Operationally, this saving is comparable to the link-budget difference between adjacent SFs (order of a few $\mathrm{dB}$), and thus meaningful for airtime/energy trade-offs \cite{chaudhariLPWANTechnologiesIoT}. A fixed $10\,\mathrm{dB}$ heuristic undershoots in this environment (held-out achieved reliability $\hat{\rho}\approx 0.923$--$0.947$, computed on received packets), motivating data-driven margins for reliability planning.} Note that all fade margins here are calibrated from residuals of models that include SNR. If SNR is omitted at planning time, the same pipeline applies, but the calibrated $M_{\mathrm{F}}(p)$ increases slightly, an ablation that we treat as part of our future work.

\subsection{Discussion}
\label{subsec:discussion}

\CHANGES{A summary of answers to our research questions (\textbf{RQ1}--\textbf{RQ4}) (listed in Table~\ref{tab:RQS}) is as follows. First, with respect to \textbf{RQ1}, environment-aware covariates and the receiver-reported SNR contribute beyond geometry and structure. HC3 \textit{Type}~II/III ANOVA yields a coherent effect-size ordering in which distance and wall losses dominate (e.g., partial $\eta^2\approx 0.21$ for $n$ and $\approx 0.16$ for $L_{\mathrm{brick}}$), while SNR remains substantial (partial $\eta^2\approx 0.12$). Environmental terms are individually smaller yet directionally consistent. Nested block tests further confirm that adding the environmental block after structure and adding SNR afterward each deliver large, non-redundant gains (Table~\ref{tab:mlr_significance}). Second, addressing \textbf{RQ2}, model comparison under the same chronological $80{:}20$ hold-out and $5$-fold time-ordered cross-validation shows that linear MLR (including regularized variants) and BLR are effectively tied in point accuracy, whereas the selective second-order polynomial applied only to continuous predictors provides the best bias--variance trade-off, improving generalization (RMSE from $\approx 8.24\,\mathrm{dB}\ \text{to}\ 7.38\,\mathrm{dB}$ and $R^2$ from $0.81$ to $0.84$, with consistent held-out performance; Table~\ref{tab:perf_comparison}). Third, in response to \textbf{RQ3}, OOF residuals depart strongly from Gaussianity, and a compact $K{=}3$ Gaussian mixture yields the best fit among the tested families by BIC/KS (Table~\ref{tab:distfits}, Fig.~\ref{fig:qqplots}), consistent with heterogeneous indoor regimes. Finally, for \textbf{RQ4}, converting residual tails into a deployment control, prescribing fade margins from upper-tail OOF quantiles calibrates on unseen data, and at a $1\%$ outage target ($99\%$ reliability), the selective quadratic mean reduces the required margin by approximately $2.1$--$2.3\,\mathrm{dB}$ relative to linear baselines (Table~\ref{tab:fm_calibration}, Fig.~\ref{fig:rho_vs_fm}), translating improved mean modeling and residual characterization into a concrete link-budget reduction.}

\section{Conclusion}
\label{sec:conclusion}

This work develops an environment-aware indoor LoRaWAN path loss workflow that stays interpretable (distance and wall losses remain explicit) while accounting for indoor non-stationarity using measured environmental covariates and a receiver-side link indicator (SNR). Using a $12$-month office deployment at $868\,\mathrm{MHz}$, we evaluate linear MLR (with regularized variants), Bayesian linear regression, and a selective second-order polynomial extension applied only to continuous predictors, under a chronological $80{:}20$ hold-out and $5$-fold time-ordered cross-validation with within-fold preprocessing. The selective polynomial model improves generalization relative to linear baselines (cross-validated RMSE drops from about $8.24\,\mathrm{dB}$ to about $7.38\,\mathrm{dB}$ and cross-validated $R^2$ rises from about $0.81$ to about $0.84$, with consistent held-out performance). OOF residual diagnostics show clear departures from Gaussianity, and a compact three-component Gaussian mixture captures the sharp core plus a light broad tail. Turning this residual behavior into a deployment control, we prescribe fade margins from the upper-tail quantiles of OOF residuals and validate on the held-out test set. At a $1\%$ outage target, the polynomial model requires about $25.7\,\mathrm{dB}$ versus roughly $27.8$--$28.1\,\mathrm{dB}$ for linear baselines, giving an approximately $2.3\,\mathrm{dB}$ margin saving at fixed reliability.

\section{Limitations and Future Directions}
\label{sec:future}

\CHANGES{This study is site-calibrated: measurements were collected on a single floor within one building, using six fixed end devices to the gateway links, so fitted wall-loss parameters and regression coefficients should be interpreted as deployment-specific rather than universally transferable. Environmental covariates are co-located with the end devices and serve as practical proxies for indoor state, and thus their generalization across building classes, sensor placements, and HVAC/occupancy regimes remains to be established. Finally, path loss is reconstructed from gateway-reported received-power metadata for successfully received packets. Therefore, packets that are not received do not provide received-power metadata and are therefore not directly represented in the path loss residual analysis or in the tail-based margin calibration.}

Next, we will prioritize external validity and measurable state context rather than adding more model families. We will replicate the campaign across multiple buildings and floor-plan types, then use hierarchical (multi-level) analyses to separate within-site effects from between-site variability in wall losses and environmental coefficients, enabling portable priors. We will also link residual regimes to observable building state (e.g., HVAC telemetry and coarse occupancy proxies) instead of treating mixture components as purely phenomenological. On the deployment side, we will extend the same pipeline to planning-mode operation, where receiver-side link indicators may be unavailable, by refitting without such terms and recalibrating fade margins. We will then integrate the calibrated margins into LoRaWAN ADR and transmit-power control to quantify reliability–energy trade-offs, and evaluate lightweight geometry-aware features from floor plans (e.g., corridor-path descriptors) that reduce tail risk without sacrificing interpretability. Finally, we will standardize reporting via shared splits, consistent residual-law selection rules, and bootstrap uncertainty for tail quantiles to support fair, reproducible comparisons across indoor LoRaWAN studies.

\backmatter

\section*{Abbreviations and Acronyms}

\newlist{abbrv}{description}{1}
\setlist[abbrv]{
  style=multiline,
  font=\normalfont\mdseries, 
  labelwidth=72pt,          
  leftmargin=!,
  labelsep=0.8em,
  itemsep=0.2ex, parsep=0pt, topsep=0.2ex
}
\makeatletter
\renewcommand{\descriptionlabel}[1]{%
  \hspace\labelsep\makebox[72pt][l]{\normalfont\mdseries #1}}
\makeatother

\noindent\small
\begin{abbrv}
\item[AIC] Akaike Information Criterion
\item[ANOVA] Analysis of Variance
\item[BCa] Bias-Corrected and Accelerated (bootstrap)
\item[BIC] Bayesian Information Criterion
\item[BLR] Bayesian Linear Regression
\item[CI] Confidence Interval
\item[df] Degrees of freedom
\item[GMM] Gaussian Mixture Model
\item[HC3] Heteroskedasticity-consistent covariance estimator (MacKinnon--White HC3)
\item[HVAC] Heating, Ventilation, and Air Conditioning
\item[IoT] Internet of Things
\item[KDE] Kernel Density Estimation
\item[KS] Kolmogorov--Smirnov (test)
\item[Lasso] Least Absolute Shrinkage and Selection Operator
\item[LoRaWAN] Long-Range Wide Area Network
\item[LoS] Line-of-Sight
\item[MLR] Multiple Linear Regression
\item[MWM] Multi-wall Model
\item[NIG] Normal--Inverse-Gamma (prior)
\item[NLoS] Non-Line-of-Sight
\item[OOF] Out-of-Fold
\item[OLS] Ordinary Least Squares
\item[POLY2] Selective second-order polynomial mean model (continuous predictors only)
\item[RMSE] Root Mean Square Error
\item[RSSI] Received Signal Strength Indicator
\item[SF] Spreading Factor
\item[SNR] Signal-to-Noise Ratio
\end{abbrv}
\normalsize


\section*{Declarations}

\subsection*{Availability of data and materials}
The dataset and analysis scripts supporting this study are publicly available at:
\url{https://github.com/nahshonmokua/LoRaWAN-Indoor-PL-parametrics}.
The repository includes code to reproduce the data preparation, model fitting, cross‑validation, residual diagnostics, and fade‑margin calibration reported in the manuscript. Any additional information is available from the corresponding author upon reasonable request.

\subsection*{Conflict of interest}
The authors declare that they have no conflict of interest.

\subsection*{Funding}
This work was partly supported by the German Academic Exchange Service (DAAD) through the Kenyan–German Postgraduate Training Programme 2023/2024 under Grant 57652455. Additional funding was provided by the Deutsche Forschungsgemeinschaft (DFG, German Research Foundation) under Grant 425868829 as part of Priority Program SPP2199: Scalable Interaction Paradigms for Pervasive Computing Environments.

\subsection*{Authors' contributions}
\textbf{N.M.O.} designed and executed the measurement campaign; implemented data ingestion, cleaning, and feature engineering; developed and evaluated the regression models (MLR, regularized variants, Bayesian linear regression, and second‑order polynomial); performed ANOVA and residual diagnostics; carried out the fade‑margin calibration; prepared figures and tables; and drafted the manuscript. \textbf{K.V.L.} conceived the study; supervised the methodology and experimental design; contributed to the interpretation of results and to writing and critical revision of the manuscript. All authors read and approved the final manuscript.

\subsection*{Acknowledgements}
The authors thank the Ubiquitous Computing Group at the University of Siegen for support with the indoor deployment and infrastructure, and for helpful discussions during the year‑long campaign.

\subsection*{Authors' information}
Not applicable.

\subsection*{Ethics approval and consent to participate}
Not applicable. The study involves environmental sensing and wireless link measurements. No personal or identifiable data, headcounts, or movement traces were collected.

\subsection*{Consent for publication}
Not applicable.

\subsection*{Materials availability}
Not applicable. No new unique biological or chemical materials were generated. Commercial, off‑the‑shelf hardware is listed in the manuscript; build details can be provided on request.

\subsection*{Code availability}
The analysis and plotting code are available in the same public repository listed above under “Availability of data and materials”.

\bibliography{references}

\end{document}